\documentclass[usenatbib,useAMS,journal]{rmaa}

\usepackage{paralist}

\usepackage{psfrag,color}

\usepackage[latin1]{inputenc}

\title{Chemical Compositions of RV Tauri Stars and Related Objects} 

\author{
  S. Sumangala Rao\altaffilmark{1} 
  and Sunetra Giridhar\altaffilmark{1}}

\altaffiltext{1}{Indian Institute of Astrophysics, Bangalore,
  India.}
\shortauthor{Rao \& Giridhar}
\shorttitle{Chemical Compositions of RV Tauri Stars and Related Objects}
\fulladdresses{
\item S. Sumangala Rao and Sunetra Giridhar: Indian Institute of Astrophysics, 
 Bangalore 560034, India. 
(sumangala, sunetra@iiap.res.in).}
\listofauthors{S. Sumangala Rao \& Sunetra Giridhar}
\indexauthor{Rao, S. S.}
\indexauthor{Giridhar, S.}

\abstract{
 We have undertaken a comprehensive abundance analysis for a sample of relatively
unexplored RV Tauri and RV Tauri like stars to further our understanding of
post-Asymptotic Giant Branch (post-AGB) evolution. 
From our study based on high resolution spectra and grid of model atmospheres,
 we find indications of mild s-processing for V820 Cen and IRAS 06165+3158. On the other hand, SU Gem and BT Lac exhibit the effects of mild dust-gas winnowing.
We have also compiled the existing abundance data on RV Tauri objects and find
that a large fraction of them are afflicted by dust-gas winnowing and
 now added by the present work, we find a small
group of two RV Tauris showing mild s-process enhancement in our Galaxy. With two
out of three reported s-process enhanced objects belonging to RV Tauri spectroscopic
class C, these intrinsically metal-poor objects appear to be promising candidates
to analyse the possible s-processing in RV Tauri stars.
}

\resumen{}


\begin{document}
\maketitle

\section{Introduction}
\label{sec:intro}
RV Tauri stars are pulsating variables located in the instability
strip along with the Cepheids but at relatively lower luminosities.
Their characteristic light curves show alternating deep
 and shallow minima with a formal period
(time elapsed between two consecutive deep minima) of 30-150 days.
Photometrically there are two types of RV Tauri stars, RVa and RVb
 (Kukarkin, Parenago \&  Kholopov 1958): RVa stars show a
 quasi-constant brightness of mean    
 light whereas RVb stars exhibit a longer term variation
in mean brightness with a period of about 600-1500 days.
  Waelkens \& Waters (1993) proposed that the RVb phenomenon is
 caused by either the dust ejected by the star or obscuration by
 the circumstellar shell as the star moves
 in the binary orbit. Extended multicolor photometry of
 a large sample of RV Tauri stars by
 Pollard et al. (1996) showed further complexity such as damping of the
 short term (pulsational) variations at long term minima for a few RV Tauri stars
 which was attributed to an interaction with the previously ejected matter or
 with the companion during certain orbital phase thereby affecting the pulsation.
 The absence of secular variations in RVa does not necessarily imply that they are single stars; in fact,
 RVa objects are known to have binary companions e.g. AC Her
 (see Van Winckel et al. 1998) and RU Cen (see Maas, Van Winckel \& Waelkens 2002).
 From a comprehensive study of RV Tauri binaries, Van Winckel et al. (1999);
 Maas, Van Winckel \& Waelkens (2002) proposed that    
 the RV Tauri photometric types do
  not arise from a physical difference like Spectral Energy Distributions (SEDs) and chemical composition 
 but mainly from the viewing angle onto the disk. In this scenario the
  RVa objects are thought to be the ones with low inclination while objects
 with high inclination such that the disk is seen edge-on would appear as RVbs.

Preston et al. (1963) classified RV Tauri variables spectroscopically
into RVA, RVB and RVC.
RVA have spectral type G-K and near light minimum show
TiO bands of abnormal strength.
RVB are relatively warm weak-lined objects of
spectral type F and exhibit strong CN and CH bands at light minimum.
RVCs have weak metal lines in their spectra and have high radial velocities (Joy 1952).
The CN and CH bands are weaker or absent at all phases. They are genuinely metal-poor
objects.

 As suggested by Wallerstein (2002) and Van Winckel (2003) in their reviews
 RV Tauri stars are post-AGB objects crossing the instability strip.
 With the detection of RV Tauri stars in the Large Magellanic Cloud (LMC) by Alcock et al. (1998),
their location on the high luminosity end of the population II
instability strip is confirmed.  Using the known distance modulus of
the LMC, an absolute magnitude of $-$4.5 was estimated for RV Tauri variables with
fundamental periods (time elapsed between a deep and a shallow minima) of about 50
days using the calibrated P-L-C (Period-Luminosity-Color) relation by Alcock et al. (1998)
further supporting the above suggestion.
The detection of IR fluxes (Jura 1986) and the  high estimated luminosities
(Alcock et al. 1998) supports the idea that these stars are in the post-AGB phase
evolving towards the blue in the Hertzsprung-Russell (H-R) diagram.

Studies of the chemical compositions of RV Tauri variables were undertaken initially
in large part to glean information about their evolutionary status and in
particular about the compositional changes wrought by internal nucleosynthesis
and mixing processes (Dredge-ups).
 However, a considerable fraction of them exhibited a very different abundance peculiarity$-$a
systematic depletion of refractory elements.
A strong signature of this phenomenon has been
observed in the post-AGB objects like HR 4049, HD 52961, BD+39$^{o}$4926, HD 44179
etc (see Van Winckel 2003 for a review). Through their study of $\lambda$ Bootis
stars showing similar depletions, Venn \& Lambert (1990) noted the
resemblance of the observed abundance pattern  with
that of the interstellar gas in which the metals are
depleted through fractionation in the interstellar grains.  
Bond (1991) suggested that the extreme metal deficiency of HR 4049 like objects
could be caused by the selective removal of metals through grain formation. 
A semi-quantitative model to explain this phenomenon observed in $\lambda$ Bootis stars
 and HR 4049 like objects was developed by Mathis \& 
  Lamers (1992). These authors proposed two scenarios: capture by 
  the presently visible post-AGB star of the depleted gas from the binary companion
  or rapid termination of a vigorous stellar wind in a single star so that grains
  are blown outwards (and hence lost) resulting in a  photosphere  devoid of
  these grain forming elements. Waters, Trams \& Waelkens (1992) proposed an
  alternate scenario based upon slow accretion from the circumstellar or circum-system disk.
  This scheme provides favourable conditions for this effect to operate without any 
  restriction on the nature of the binary companion. 
More observational support of this hypothesis such as large [Zn/Fe]
for HD 52961 (Van Winckel, Mathis \& Waelkens 1992) and strong correlation between
stellar abundance for IW Car and depletions observed in the interstellar gas 
demonstrated by Giridhar, Rao \& Lambert (1994) resulted in further 
detections of RV Tauri and post-AGB objects 
showing this effect commonly referred as 'dust-gas winnowing' or 'dust-gas separation'
 A summary of these detections can be found in recent papers like
Sumangala Rao, Giridhar \& Lambert 2012, Van Winckel et al. 2012.   
The condensation temperature (T$_C$)\footnote{The condensation temperature T$_C$ is
the temperature
at which half of a particular element in a gaseous environment condenses into dust
grains.}
 being an important parameter measuring 
the propensity of a given element into grain formation, the dependence of
the observed abundance on T$_C$ can be used to identify these objects.
 For brevity, hereinafter we would refer to the
 'dust-gas winnowing effect' as DG effect.

Among RV Tauris, this effect is most prevalent in RVB objects while their
cooler sibling RVA only show weak manifestations possibly due to the dilution
caused by their deep convective envelopes. The genuinely metal-poor RVCs
are unaffected by the DG winnowing since their metal-poor environment 
are not conducive for grain formation (Giridhar, Lambert \& Gonzalez 2000).

In the present work we have enlarged the RV Tauri sample by studying six
unexplored RV Tauri
stars and IRAS\footnote{Infrared Astronomical Satellite} objects located in or near the RV Tauri box in the IRAS two color
diagram.
We also present a more recent abundance analyses for the RVC star V453 Oph and the extremely
 depleted star HD 52961 which exhibits RVb like phenomenon in its light curve.

\section{Selection of the sample}
\label{sec:sample selection}

Our sample (see Table~\ref{table1}) comprises mainly of unexplored RV Tauri stars,
 objects having RV Tauri like IR colors. We have also studied known RV Tauri object V453 Oph
 and heavily depleted object HD 52961 for which we provide a contemporary analysis covering more elements. 
 The IR fluxes of known RV Tauri stars have been
investigated by Lloyd Evans (1985) and Raveendran (1989). Lloyd Evans (1999) 
reported that the RV Tauri stars fall in a well-defined region of the IRAS two-color
 diagram called the RV Tauri box. This box is defined from the observed properties
 of RV Tauri dusty shells such as the temperatures at the inner boundary of
 the dust shell (T$_{O}$) and the absorption coefficient (Q) which depends
 on the density and the temperature distribution of the dust as well as on
 the chemical composition and the physical properties (like size) of the
 dust grains. Raveendran (1989) from his study of 17 sample RV Tauri stars
 found that T$_{O}$ had a range between 400-600K and Q between 0.15 to 0.5. 

The study of SEDs of six RV Tauri objects by De Ruyter et al. (2005)
 showed a large near IR excess but low line of sight extinction, this  coupled with energy balance
 considerations suggested that likely distribution of the circumstellar dust is that
 of a dusty disk. Lloyd Evans (1999) suggested that RV Tauris are those stars
 with dusty disks which are currently located within the instability strip.
Lloyd Evans (1999) hence proposed that
 this RV Tauri box in the IRAS [12]$-$[25], [25]$-$[60] diagram enclosed by
the limits [12]$-$[25]=1.0$-$1.5 and [25]$-$[60]=0.20$-$1.0 when supplemented by
large near IR flux
provides an alternative method of searching for RV Tauri stars among IRAS objects.

In fact, the photometric monitoring of IRAS sources following the above mentioned
 criteria did result in finding new samples of RV Tauri objects 
studied by Maas, Van Winckel \& Waelkens (2002) and Mass, Van Winckel \& Lloyd Evans (2005).

\begin{figure}
\begin{center}
\includegraphics[width=8.5cm,height=9cm]{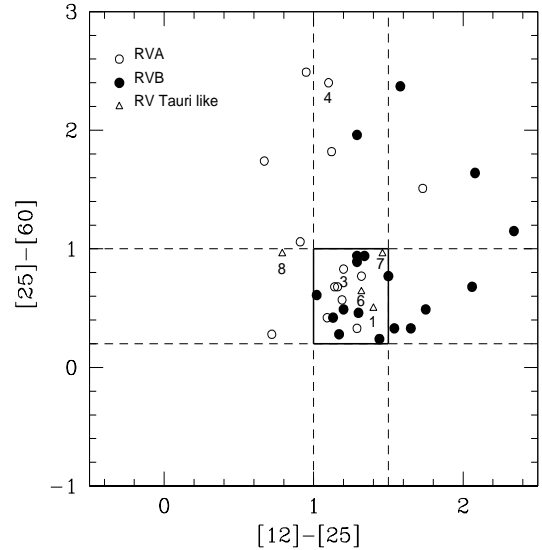}
\caption{The IRAS color-color diagram showing the "RV Tauri" box.
 The figure contains our sample stars and all the well studied
RV Tauri stars. The program stars are numbered according to Table~\ref{table1}.
}
\label{loci1}
\end{center}
\end{figure}

In Figure \ref {loci1} we have plotted our program stars (those having IR colors)
which have been numbered according to Table~\ref{table1}. In the figure we have also
plotted RV Tauri stars with known spectroscopic classification.
Most of our program stars with the exception of BT Lac are located
in or around the RV Tauri box.

Although a fraction of known
RV Tauri stars are found in RV Tauri Box, many well known RV Tauri
stars do not conform to these limits and lie outside the box. 
Perhaps the limits of RV Tauri box  needs upward revision in both axes.
Nevertheless this box provides a starting point 
 for identifying RV Tauri candidates among IRAS
sources with no photometry. In what follows we will refer as "RV Tauri like"
those objects with RV Tauri like colors in IRAS two colour diagram without
photometric confirmation.

\begin{figure}
\begin{center}
\includegraphics[width=8cm,height=7cm]{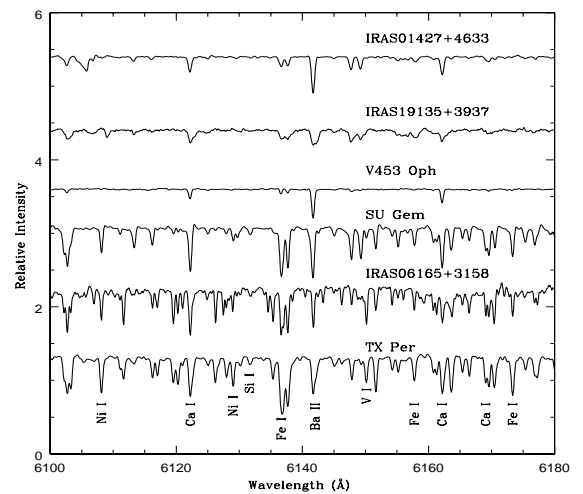}
\caption{Sample spectra of our program stars presented in the descending order of
temperature (top to bottom) in the 6100-6180\AA~ region.}
\label{loci2}
\end{center}
\end{figure}

\begin{table*}
 \centering
 \begin{minipage}{140mm}
  \caption{The Program Stars.}
 \label{table1}
\begin{tabular}{lllll}
  \hline
\multicolumn{1}{l}{No.}&
\multicolumn{1}{l}{IRAS}&
\multicolumn{1}{l}{Other Names}&
\multicolumn{1}{l}{Period (Days)}&
\multicolumn{1}{l}{Var Type} \\

 \hline
1& 06165+3158 & ...& ...& ...\\
2&  ...&V820 Cen, SAO 205326& 150& RV Tauri  \\
3& 06108+2743&  SU Gem, HD 42806& 50& RV Tauri \\
4& 22223+5556& BT Lac& 41& RV Tauri \\
5&       ...  &  TX Per& 78& RV Tauri \\
6& 19135+3937  &...& ...& ...\\
7& 01427+4633& SAO 37487, BD+46$^{o}$442& ...&... \\
8& 07008+1050& HD 52961, PS Gem& 71& SRD$^{\ast}$ \\
9&      ...  &  V453 Oph, BD-02$^{o}$4354& 81& RV Tauri \\
\hline
\end{tabular}
\flushleft$^{\ast}$ {SRDs are semi-regular variable giants and supergiants of spectral types F, G and K.
 Sometimes emission lines are seen in their spectra. They have pulsation periods in the range of
 30-1100 days with an amplitude of variation upto the 4$^{th}$ magnitude in their light curves.}
\end{minipage}
\end{table*}

\begin{table*}[!t]
\centering
\caption{Stellar Parameters Derived from the
Fe-line Analyses}
\setlength{\tabnotewidth}{0.95\linewidth}
\setlength{\tabcolsep}{0.7\tabcolsep} \tablecols{10}
\label{table2}
\begin{tabular}{lllcrlcrlr}
 \hline
\multicolumn{1}{l}{Star} &
\multicolumn{1}{l}{UT Date}&
\multicolumn{1}{l}{V$_{r}$\tabnotemark{a}}&
\multicolumn{1}{c}{T$_{\rm eff}$, $\log g$,
[Fe/H]} &
 \multicolumn{1}{c}{$\xi^{\rm b}_{\rm t}$} &
 \multicolumn{2}{c}{Fe
I\tabnotemark{c}} & &
\multicolumn{2}{c}{Fe II\tabnotemark{c}}   \\ \cline{6-7} \cline{9-10}
&& \multicolumn{1}{l}{(km~s$^{-1}$)} &&
\multicolumn{1}{l}{(km~s$^{-1}$)} &
\multicolumn{1}{l}{$\log \epsilon$}& \multicolumn{1}{l}{n} & & \multicolumn{1}{l}{$\log \epsilon$} &\multicolumn{1}{l}{n} \\
 \hline
IRAS 06165+3158 & 2007 Nov 5&$-$16.0& 4250,1.50, $-$0.93 & 2.8\phantom{000}&$6.54 \pm 0.15$ &
55 && $6.49 \pm 0.17$  &6\\
V820 Cen& 2011 Mar 02&$+$259.0& 4750, 1.5, $-$2.28& 2.4\phantom{000} &$ 5.14\pm0.16$& 67&& $5.21\pm 0.15$ &8 \\
V820 Cen& 2011 Mar 03&$+$266.0 & 4750, 1.5, $-$2.35 & 2.4\phantom{000} &$ 5.06\pm0.17 $& 39&& $5.15\pm 0.01$ &2 \\
IRAS 06108+2743 & 2009 Dec 26&$+$8.9 & 5250, 1.00 ,$-$0.25 & 3.0\phantom{000} & $7.20 \pm 0.10$ &
31 && $7.19 \pm 0.10$ & 14 \\
IRAS 22223+5556& 2009 Oct 10&$-$70.0& 5000, 2.00, $-$0.17 & 3.7\phantom{000}& $7.28\pm 0.13$ & 36 && $7.27\pm 0.14$ & 4 \\
TX Per & 2009 Dec 27&$-$17.0& 4250,1.50, $-$0.58 & 3.1\phantom{000} & $6.81 \pm 0.13$ &
62 && $6.94 \pm 0.13$  &7  \\
IRAS 19135+3937&2007 Nov 3&$-$13.2& 6000,0.50,$-$1.04 & 4.1\phantom{000}& $6.46\pm 0.15$ &
26 && $6.36 \pm 0.09$ &  7 \\
IRAS 01427+4633&2007 Dec 21&$-$98.2& 6500,0.50,$-$0.79 & 3.5\phantom{000}& $6.71\pm 0.09$ &
37 && $6.61 \pm 0.11$ & 8 \\
HD 52961 & 2011 Jan 27&$+$4.2 & 6000, 0.5, $-$4.55 & 5.1\phantom{000} &$ 2.91\pm
 0.00$ & 1 && $2.89\pm 0.07$ &2 \\
V453 Oph & 2009 May 10&$-$125.9& 5750,1.50, $-$2.26 & 3.7\phantom{000} & $5.19 \pm 0.10$ &
28 && $5.18 \pm 0.11$  &7 \\
\hline
\end{tabular}
\flushleft$^{a}${V$_{r}$ is the radial velocity in {km~s$^{-1}$}}, $^{b}${$\xi_{\rm t}$ is the microturbulence}
\flushleft$^{c}${$\log \epsilon$ is the mean abundance relative to H
(with $\log \epsilon_{\rm H} = 12.00$).
The standard deviations of the means as calculated
from the line-to-line scatter are given.
$n$ is the number of
considered lines.}
\end{table*}

\section{Observations}
\label{sec:observations}
 
High-resolution optical spectra were obtained at the W.J. McDonald
Observatory with the 2.7m Harlan J.
 Smith reflector and the Tull coud\'e spectrograph (Tull et al. 1995).  This spectrometer gives a
resolving power of about 60,000 and a broad spectral range was covered in a
single exposure. A S/N ratio of 80-100 over much of the spectral range was achieved.
Figure \ref{loci2} illustrates the resolution and quality of sample spectra of our program stars in the
wavelength region 6100-6180\AA~.
The sample spectra have been arranged in the order of decreasing effective
temperatures.

Program stars IRAS 06165+3158 and TX Per have the same effective temperature.
The spectra of HD 52961 and V820 Cen were obtained with the echelle spectrometer of
the 2.34m Vainu Bappu Telescope at the Vainu Bappu Observatory (VBO) in Kavalur, India
giving a resolution of about 28,000 in the slitless mode (Rao et al. 2005).

\section{Abundance analysis}
\label{sec:abundance}

The method of abundance analysis and the sources of log {\itshape gf} values have
been described in
detail by  Sumangala Rao, Giridhar \& Lambert (2012), hereinafter will be referred as Paper-1.
The microturbulence velocity has been estimated by requiring that the derived
abundance are independent of the line strengths. We have used Fe \,{\sc ii} lines for warmer members
but for cooler members, the paucity of usable Fe  \,{\sc ii} lines compelled us to use Fe  \,{\sc i} lines.
The  temperatures have been determined by
demanding that the iron abundance be independent of the lower
excitation potential (LEP)
and  {\itshape  log g} have been determined from
the excitation
and ionization balance between Fe\,{\sc i} and Fe\,{\sc ii}, Ti\,{\sc i} and Ti\,{\sc ii},
 Cr\,{\sc i} and Cr\,{\sc ii}.
We could not use Hydrogen line profiles due to the presence of emission components
and asymmetries present in most of the spectra.
The derived stellar parameters have been presented in Table 2.
The sensitivity of the derived abundances to the
uncertainties of atmospheric parameters {\itshape T$_{\rm eff}$, log g} and $\xi$
are presented in Table 3. For three stars representing
the full temperature range of our sample, we present changes in
[X/Fe] caused by varying atmospheric parameters
by 200 K, 0.25 cm s$^{-2}$ and 0.5 km s$^{-1}$ (average
accuracies of these parameters) with respect to the chosen model
for each star.

\begin{table*}[t]
\centering
\caption{Sensitivity of [X/Fe] to the uncertainties\\
in the model parameters for a range of \\
 temperatures covering our sample stars.}
\setlength{\tabnotewidth}{0.5\linewidth}
\setlength{\tabcolsep}{0.8\tabcolsep}
\label{table3}
\begin{tabular}{llcrlcrlcr}
 \hline
&\multicolumn{3}{c}{TX Per}&
\multicolumn{3}{c}{IRAS 06108+2743}&
\multicolumn{3}{c}{IRAS 01427+4633} \\
\cline{2-10}
&& \multicolumn{1}{l}{(4250K)}
&&& \multicolumn{1}{l}{(5250K)}
&&& \multicolumn{1}{l}{(6500K)} \\
\multicolumn{1}{l}{Species} &
\multicolumn{1}{c}{$\Delta$$T_{\rm eff}$}&
\multicolumn{1}{c}{$\Delta$log~$g$}&
\multicolumn{1}{c}{$\Delta \xi$}&
\multicolumn{1}{c}{$\Delta$$T_{\rm eff}$}&
\multicolumn{1}{c}{$\Delta$log~$g$}&
\multicolumn{1}{c}{$\Delta \xi$}&
\multicolumn{1}{c}{$\Delta$$T_{\rm eff}$}&
\multicolumn{1}{c}{$\Delta$log~$g$}&
\multicolumn{1}{c}{$\Delta \xi$} \\
& \multicolumn{1}{c}{$-200K$}&
\multicolumn{1}{c}{$+0.25$}&
\multicolumn{1}{c}{$+0.5$}&
\multicolumn{1}{c}{$-200K$}&
\multicolumn{1}{c}{$+0.25$}&
\multicolumn{1}{c}{$+0.5$}&
\multicolumn{1}{c}{$-200K$}&
\multicolumn{1}{c}{$+0.25$}&
\multicolumn{1}{c}{$+0.5$} \\
\hline
C I & ...& ...& ...& $-$0.29& $-$0.07& $-$0.09 & $-$0.08& $-$0.01& $-$0.05  \\
N I & ...& ...& ...& ...& ...& ... & $-$0.13& $-$0.06& $-$0.04  \\
O I& $+$0.24& $-$0.01& $-$0.10& $+$0.03& $-$0.06& $-$0.09& ...& ...& ...  \\
Na I& $+$0.41& $+$0.22& $-$0.08& $+$0.05& $+$0.06& $-$0.07& ...& ...& ... \\
Mg I& $+$0.25& $+$0.15& $-$0.02& $+$0.02& $+$0.06& $-$0.04& $+$0.03& $+$0.05& $-$0.01 \\
Al I& $+$0.37& $+$0.19& $-$0.11 & $+$0.03& $+$0.05& $-$0.10& $+$0.00& $+$0.06& $-$0.05  \\
Si I& $-$0.05& $+$0.05& $-$0.08& $+$0.02& $+$0.05& $-$0.10& $+$0.00& $+$0.06& $-$0.06 \\
Si II& ...& ...& ...& ...& ...& ...& $-$0.10&$-$0.07&$+$0.06 \\
S I& ...&...&...& $+$0.24& $-$0.06& $-$0.09&$-$0.05& $+$0.03& $-$0.06 \\
Ca I& $+$0.48& $+$0.23& $-$0.01& $+$0.09& $+$0.06& $-$0.01& $+$0.04& $+$0.07& $+$0.00 \\
Sc II& ...& ...& ...& $-$0.02& $-$0.06& $-$0.01& $-$0.01& $-$0.05& $-$0.04  \\
Ti I& ...& ...& ...& $-$0.18& $+$0.07& $-$0.08& $+$0.07& $+$0.06& $-$0.03 \\
Ti II& $+$0.13& $+$0.00& $+$0.02& $-$0.03& $-$0.06& $+$0.14& $+$0.13& $-$0.06& $+$0.14 \\
Cr I& $+$0.43& $+$0.18&$-$0.08& $+$0.18& $+$0.07& $+$0.08& $-$0.05& $+$0.06& $+$0.05 \\
Cr II& $-$0.09& $-$0.04& $-$0.06& $-$0.13& $-$0.07& $-$0.05& $-$0.07& $-$0.05& $-$0.01 \\
Mn I& $+$0.37& $+$0.21& $-$0.05& $+$0.12& $+$0.07&$+$0.00& $+$0.02& $+$0.07& $-$0.06 \\
Ni I& $+$0.19& $+$0.08& $+$0.01& $+$0.11& $+$0.06& $+$0.01& $+$0.03& $+$0.06& $-$0.04 \\
Zn I& $-$0.04& $+$0.03& $-$0.01& $+$0.03& $+$0.01& $+$0.05 & $+$0.02& $+$0.06& $-$0.05  \\
Y II& ...& ...& ...& $-$0.02& $-$0.06& $+$0.02& $-$0.02& $-$0.05& $-$0.05  \\
Ce II& $+$0.24&$-$0.01&$-$0.10& $+$0.04&$-$0.05&$-$0.07& $+$0.04& $-$0.03& $-$0.05  \\
Nd II&  ...& ...& ...& $+$0.05& $-$0.06&$-$0.10 & ...& ...& ... \\
Sm II& ...& ...& ...& $+$0.04&$-$0.05&$-$0.08& ...& ...& ...  \\
\hline
\end{tabular}
\end{table*}

\begin{table*}[t!]
\caption{Elemental Abundances for IRAS 06165+3158 and V820 Cen}
\label{table4}
\begin{tabular}{lllllllllllll}
\hline
&& \multicolumn{3}{c}{IRAS 06165+3158}&
\multicolumn{3}{r}{V820 Cen$^{a}$} && \multicolumn{3}{r}{V820 Cen$^{b}$}\\
\cline{3-13}  
\multicolumn{1}{l}{Species}&
\multicolumn{1}{l}{$\log \epsilon_{\odot}$}&
\multicolumn{1}{c}{[X/H]}&
\multicolumn{1}{l}{N}&
\multicolumn{1}{l}{[X/Fe]}&
&\multicolumn{1}{c}{[X/H]}&
\multicolumn{1}{l}{N}&
\multicolumn{1}{l}{[X/Fe]}
&& \multicolumn{1}{c}{[X/H]}&
\multicolumn{1}{l}{N}&
\multicolumn{1}{l}{[X/Fe]}
 \\
\hline
O I& 8.66 & $-0.49\pm0.00$& 1& $+0.44$ && $-0.82\pm0.00$&1& $+1.46$
&& ...\\
Na I& 6.17 & $-0.47\pm0.05$& 3& $+0.46$ && $-1.86\pm0.00$&1&
$+0.42$ && ... \\
Mg I& 7.53 & $-0.90\pm0.06$& 2& $+0.03$ &&  $-1.95\pm0.09$& 2&
$+0.33$ && $-2.04\pm0.00$& 1&$+0.31$ \\
Mg II& 7.53& ...&&&& ...&&&& $-2.07\pm0.00$& 1& $+0.28$ \\
Si I& 7.51 & $-0.86\pm0.10$&8&$+0.07$ && $-1.46\pm0.11$&4&
$+0.82$ && ... \\
Ca I& 6.31 & $-1.22\pm0.13$& 11 & $-0.29$ && $-1.95\pm0.12$& 9&
$+0.33$ && $-2.03\pm0.15$& 5&$+0.32$ \\
Sc II& 3.05 & $-0.95\pm0.02$& 1s$^{\ast}$ & $-0.03$ && 
$-1.78\pm0.16$&3&$+0.50$&&  $-1.83\pm0.00$& 1&
$+0.52$  \\
Ti I& 4.90 & $-0.96\pm0.12$& 15& $-0.03$ & & $-1.77\pm 0.06 $& 2& $+0.51$ &&
$-1.76\pm0.12$&10&$+0.59$ \\
Ti II& 4.90 & $-1.10\pm0.14$& 6& $-0.17$&& $-1.90\pm0.14$& 11&
$+0.38$ && $-1.83\pm0.13$& 3& $+0.52$ \\
Cr I& 5.64 & $-0.66\pm0.08$& 8&$+0.27$&& $-2.50\pm0.08$& 6&
$-0.22$ && $-2.45\pm0.11$&4&$-0.10$  \\
Cr II& 5.64&$-0.79\pm0.00$& 1& $+0.14$&& $-2.40\pm0.00$& 1&
$-0.12$ && ... \\
Mn  I& 5.39 &$-1.22\pm0.04$& 5& $-0.29$ && $-2.46\pm 0.01 $& 2& $-0.18$&&
 ... \\
Fe  & 7.45 & $-0.93$& & &&$-2.28$ &&&& $-2.35 $ && \\
Ni  I& 6.23 & $-0.69\pm0.06$& 6& $+0.24$ && $-2.21\pm0.15$&11&
$+0.07$&& $-2.08\pm0.08$&2&$+0.27$ \\
Zn  I& 4.60 & $-0.99\pm0.01$& 2& $-0.06$ && $-1.84\pm0.20$& 2&
$+0.44$ && ... \\
Sr I& 2.92& ...&&&& $-1.77\pm 0.00 $& 1 & $+0.51$ && $-1.71\pm0.00$&1&$+0.64$  \\
Y  II& 2.21 & $-0.55\pm0.04$&2&$+0.38$ && $-2.07\pm0.13$& 1s$^{\ast}$& $+0.21$ &&
$-1.98\pm0.15$&2 &$+0.37$  \\
Zr I& 2.59& $-0.27\pm0.07$&5&$+0.66$ &&  ...&&&& ... \\
Zr I& 2.59& $-0.27\pm0.04$&1s$^{\ast}$&$+0.66$ && ...&&&& ... \\
Zr II& 2.58 & ...&&&& $-1.80\pm 0.14 $& 3 & $+0.48$ && $-1.88\pm0.00$&1 &$+0.47 $
 \\
Ba II & 2.17& ...&&&&  $-1.74\pm 0.01 $&1s$^{\ast}$&$+0.54$ && ...   \\ 
La II& 1.13& $-0.67\pm0.07$&2&$+0.26$ && $-1.76\pm 0.17 $& 2& $+0.52$ && ...  \\
Ce II& 1.58 & $-0.63\pm0.14$& 4& $+0.30$ && $-1.93\pm0.12$&3s$^{\ast}$ &$+0.35$&& $-2.07\pm0.05$&2 &$+0.28$  \\
Pr II& 0.78& $-0.37\pm0.00$& 1s$^{\ast}$& $+0.56$ && ... &&&& ... \\
Nd II& 1.45&  $-0.63\pm0.01$& 2 & $+0.30$ && $-1.98\pm0.18$& 4& $+0.30$ && $-2.17\pm0.02$&2 &$+0.20$  \\  
Nd II& 1.45& $-0.73\pm0.03$& 1s$^{\ast}$ & $+0.20$ &&  ...&&&& ... \\ 
Sm II& 1.01& $-0.67\pm0.11$& 3&$+0.26$ && $-1.85\pm0.12$& 4& $+0.43$ && $-1.79\pm0.01$&2 &$+0.56$  \\ 
\hline
\end{tabular}
\flushleft$^{\ast}$ {The number of features synthesized for each element has been indicated.}
\flushleft$^{a}$ {The abundance measurements of V820 Cen for March 2, 2011.}
\flushleft$^{b}$ {The abundance measurements of V820 Cen for March 3, 2011.}
\end{table*}

The abundances of elements for all our program stars are presented in Tables 4, 5 and 6 
respectively.

The derived abundances relative to solar abundances are presented in these tables.
The solar photospheric abundances given by Asplund, Grevesse \& Sauval (2005) have been used as
reference.
The possible systematic effects caused
by the adopted {\itshape gf} values from different sources have been described in
our Paper 1.
For elements with lines exhibiting hyperfine (hfs) and isotopic splitting,
we have employed synthetic spectra in deriving abundances.
For elements Sc and Mn, we have used
the hfs component list and their {\itshape log gf} given by
 Prochaska \& McWilliam (2000),
 for Eu (Mucciarelli et al. 2008) and for
 Ba (McWilliam 1998).
\section{Results}
\label{sec:results}

 There are several processes that affect the stellar
composition during the course of its evolution such as a)
 the initial composition of the Interstellar Medium
(ISM), b) the effect of nucleosynthesis and mixing processes such as dredge-ups, 
c) DG winnowing that affects a large number of studied RV Tauri stars
 and post-AGBs, d) for a very small group of objects
 the abundances show dependencies on their First Ionization Potential (FIP).
 In the following subsections on
individual stars, we discuss the derived abundances to infer the influence of these
effects operating in our program stars.

\subsection{\bf New sample}

\subsubsection{IRAS 06165+3158}

 Miroshnichenko et al. (2007)
gives the spectral type as K5Ib, remark that the star is `probably
metal deficient', and describe a strong IRAS infrared excess due to cold dust.
 No OH maser emission at 1612 MHz was detected for this star (Lewis, Eder \& Terzian 1990).
No photometric observations have been
reported so the variable type and period are unknown. This star has been included in our
 analysis as it had RV Tauri like colors in the two-color diagram as can be seen in Figure \ref {loci1}.

\begin{figure}
\begin{center}
\includegraphics[width=\columnwidth]{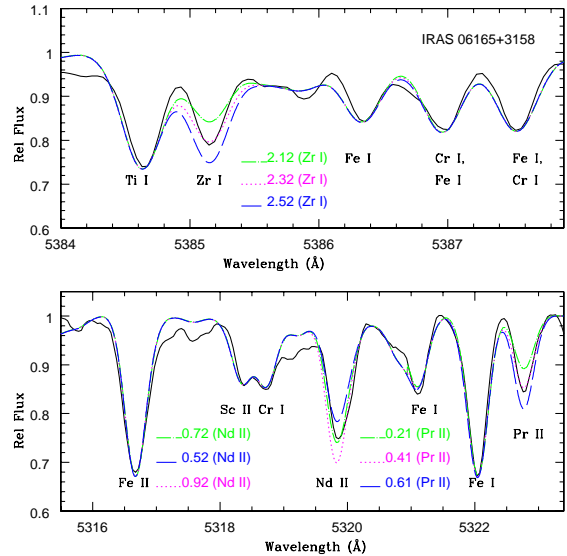}
\caption{The agreement between synthesized and observed spectrum for IRAS
06165+3158 for selected regions containing the lines of s-process elements.}
\label{loci4}
\end{center}
\end{figure}

Our abundance analysis confirms that this star is
metal-deficient ([Fe/H]  = $-$0.93). Mild enrichment of s-process elements
 (Y, Zr, La, Ce, Nd and Sm)
and also that of Pr, an r-process element
 is seen with an average [$s$/Fe] of $+$0.4 dex - see Table 4.
 Due to the low temperature of the star and very poor S/N ratio in blue the
 number of clean s-process element lines (even for synthesis) are woefully small.
 Ba\, {\sc ii} feature at 6141.7\AA~ has doubling in the core and the one at 5853.6\AA~ has a distinct unresolved component hence could not be used.
The estimated s-process abundances are supported by the synthesis of these
 features as can be seen in Figure \ref {loci4}. The O abundance has
 been determined from the forbidden line at 6300.3\AA~.
 But the C abundance is surprisingly low. The CH bands in
 the 4300\AA~ region as well as the C$_{2}$ bands in the 5150-5165\AA~ region
 appear weak and synthesis
 indicate [C/H] $<$ -2.0.

 This is atypical star in the sense that at a metallicity of $-$0.93,
 the expected enrichment of $\alpha$ elements is not seen ([Ca/Fe] = -0.3,
 [Ti/Fe] = [Si/Fe] = [Mg/Fe] = 0). Also the s-process enrichment is not
 accompanied by C-enrichment. A continuous photometric and spectrometric monitoring
 is required to detect the cause of s-process enhancement (binarity ?).

\subsubsection{V820 Cen}

V820 Cen is listed as an RV Tauri variable in the General Catalogue of Variable Stars (GCVS)
 with a period of 150
days but photometry (Eggen 1986; Pollard et al. 1996) shows considerable
variations in the light curve with Pollard et al. finding three
main periods (148, 94 and 80 days). It is assigned photometric type RVa and the
spectroscopic type has not been given.

 We had two spectra of this object observed on March 2 and 3, 2011.
 The echelle grating setting being different on these two nights,
 the coverage in each echelle order was different although some
 overlap existed. Hence we have conducted two independent abundance
 analyses for these two spectra. We found the same atmospheric parameters
 for these two epochs which is not surprising given the long period of the object.
The abundance analysis (Table 4) shows the star to be very metal-poor
([Fe/H] = $-$2.3 ) and unaffected by DG winnowing: the
$\alpha$-elements (Mg, Ca, Ti) have their expected [$\alpha$/Fe]
values and Sc is not underabundant. These abundances and the high
radial velocity ($+$265 km s$^{-1}$) suggest halo membership.
 The C\,{\sc i} lines are below the detection limit and the forbidden lines
 of O\,{\sc i} indicate enhanced O abundance.

\begin{figure}
\begin{center}
\includegraphics[width=8.5cm,height=9cm]{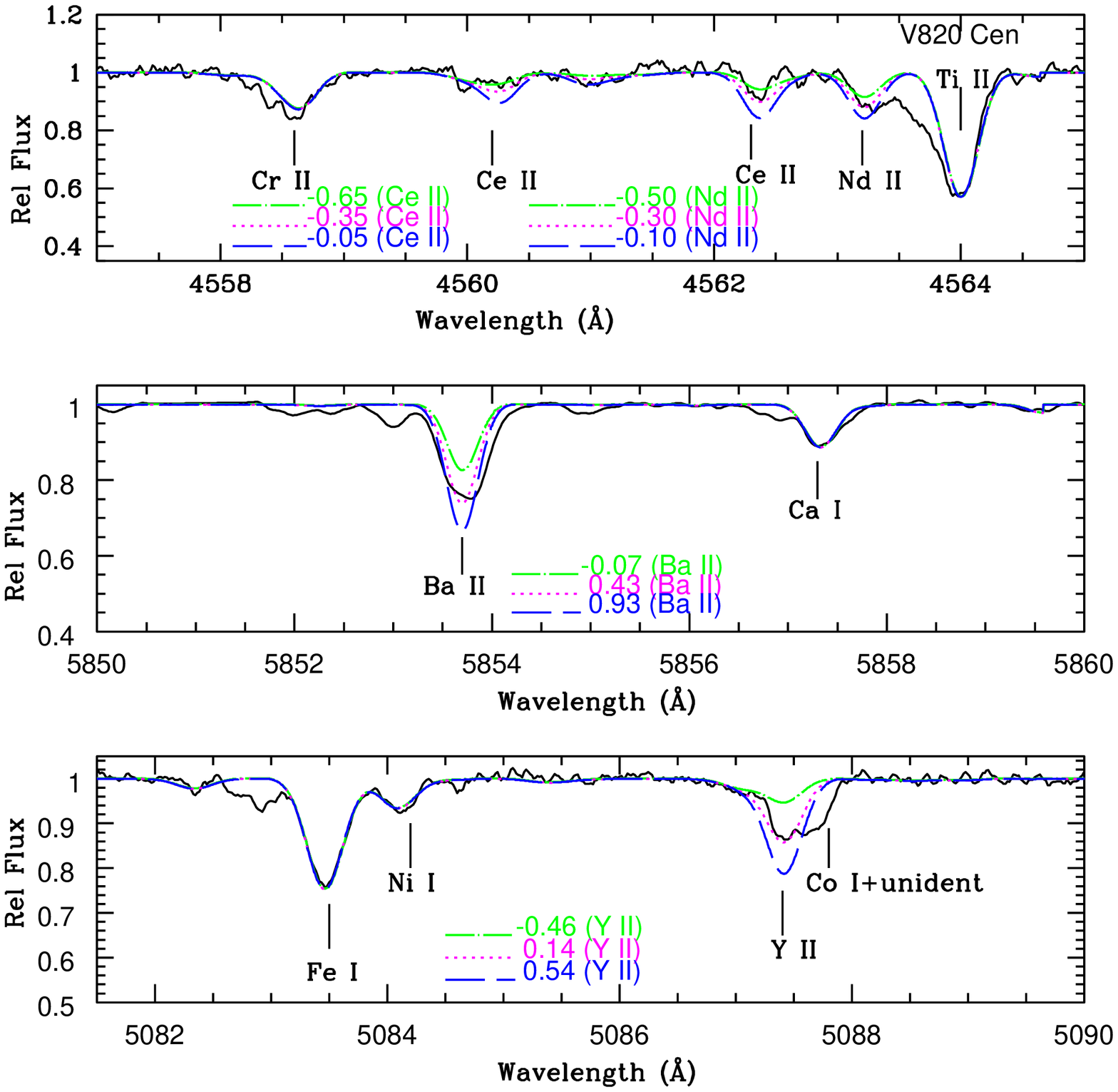}
\caption{The agreement between synthesized and observed spectrum for V820 Cen
for selected regions containing the lines of s-process elements.}
\label{loci7}
\end{center}
\end{figure}

 The most interesting feature was the detection of lines of several
 s-process elements present in both spectra.
  The metal-poor nature of this star was very helpful in detecting
  the features of these elements. In Figure \ref{loci7} we have shown agreement
  between the synthesized and observed spectrum for several s-process
  elements. We do not find significant difference between the light
  and heavy s-process elements.

 This object with RVC spectral characteristics
 however looks more or less like a twin of V453 Oph given the
 high radial velocities, low metallicity, C underabundance and mild s-process enrichment.

\subsubsection{IRAS 06108+2743}

 More popularly known as SU Gem, this star is a RVb variable with a pulsation
period of 50 days and a long period of 690 days (Joy 1952).
 Preston et al. (1963) and Lloyd Evans (1985)
 assigned the spectroscopic type RVA.
 De Ruyter et al. (2005) constructed the
 SED for SU Gem and used an optically
 thin dust model to estimate the parameters of the dust shell and
they suggested the possible dust distribution to be in a stable Keplerian disk.
 Detailed study of the IR spectra of SU Gem (Gielen et al. 2008)
 indicated the presence of amorphous and crystalline silicates pointing towards an O-rich disk.

\begin{figure}
\begin{center}
\includegraphics[width=\columnwidth]{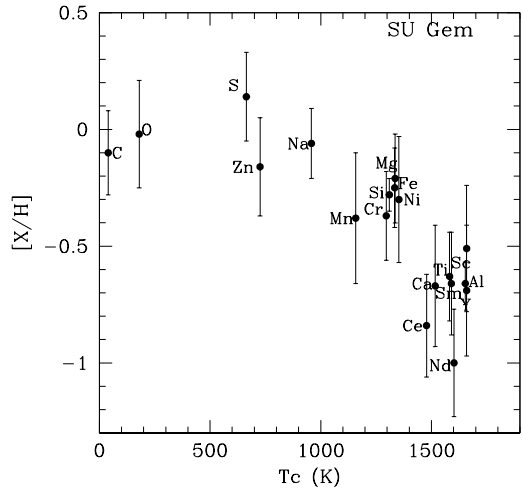}
\caption{Plot of [X/H] versus T$_C$ for SU Gem}
\label{loci3}
\end{center}
\end{figure}

 The chosen model (Table 2)
 gives the  abundances in Table 5.
 Ionization equilibrium is accounted  with
$\Delta$ = [X$_{II}$/H] $-$ [X$_{I}$/H]
 is $-0.01$, $+0.11$ and $+0.02$ for Fe, Ti and Cr respectively.
The C abundance was derived from the synthesis of the
C\,{\sc i} line at 6587\AA\ and the O abundance from the 6300 and
6363 \AA\ forbidden lines.
 It is evident from Figure \ref{loci3} that depletion of elements with the highest T$_C$s
like Ca, Sc, Ti, Al as well as the s-process elements point to
mild DG winnowing.

\subsubsection{IRAS 22223+5556}

Also known as BT Lac, this star is a RV Tauri variable with a period of 40.5 days of the RVb class with
a long period of 654 days (Tempesti 1955; Percy et al. 1997).
 The star was observed on the night of October 10, 2009 and has a
 V magnitude of 12.8 which is at the faint limit of the telescope.
 Hence the S/N ratio was only about 30-40
 even after co-adding four exposures each of 30 min duration.
 The number of usable lines were much smaller than what one
 would expect for this temperature and gravity due to distorted profiles and suggestions
 of line doubling for several elements. Our analysis relies upon clean symmetrical lines.
The metallicity of BT Lac is almost solar ([Fe/H]=$-$0.2).
 The O abundance has been determined from the forbidden line at 6300\AA~. The lines of C and S
were distorted and appeared to be double and could not be used in our analysis.
Mild DG winnowing is suggested by almost solar Zn abundance and underabundances of
Ca, Sc, Ti and the s-process elements Y, Zr, Ce and Sm $-$ see Table 5.

\subsubsection{TX Per}

This star was classified as a RVa variable in the 
GCVS (see also Percy \& Coffey 2005).
Zsoldos (1995) gave a period of 78 days.
TX Per does not have IRAS colors. Planesas et al. (1991)
 showed that TX Per had no detected OH maser emission and had
 weak CO emission pointing to a deficiency of molecules in its envelope.

The abundance analysis (see Table 5) shows that the star is mildly metal-poor
([Fe/H] = $-$0.58).
 The C and N abundances could not be determined due to the
 low temperature of the star.
The O abundance has been determined using the forbidden lines
at 6300 and 6363 \AA~.
 The high condensation temperature elements (T$_C$ $>$ 1500K ) including  $\alpha$ elements Ca, Ti
 and the s-process elements Zr and Ce again with
T$_C$ $>$ 1500K are slightly under-abundant but no signature of DG winnowing is seen as the low T$_C$ element Zn is also underabundant.

\begin{table*}[t!]
\begin{changemargin}{-1.7cm}{-1.7cm}
\caption{Elemental Abundances for IRAS 06108+2743, IRAS 22223+5556, TX Per and IRAS 19135+3937}
\setlength{\tabcolsep}{0.5\tabcolsep}
\label{table5}
\begin{tabular}{lllllllllllllllll}
\hline
&& \multicolumn{3}{c}{IRAS 06108+2743}&&
\multicolumn{3}{c}{IRAS 22223+5556} && \multicolumn{3}{c}{TX Per} && \multicolumn{3}{r}{IRAS 19135+3937}\\
\cline{3-17}
\multicolumn{1}{l}{Species}&
\multicolumn{1}{l}{$\log \epsilon_{\odot}$}&
\multicolumn{1}{c}{[X/H]}&
\multicolumn{1}{l}{N}&
\multicolumn{1}{l}{[X/Fe]}&&
\multicolumn{1}{c}{[X/H]}&
\multicolumn{1}{l}{N}&
\multicolumn{1}{l}{[X/Fe]}&&
\multicolumn{1}{c}{[X/H]}&
\multicolumn{1}{l}{N}&
\multicolumn{1}{l}{[X/Fe]}&&
\multicolumn{1}{c}{[X/H]}&
\multicolumn{1}{c}{N}&
\multicolumn{1}{c}{[X/Fe]}
 \\
\hline
C I& 8.39 & $-0.10\pm0.01$&1s$^{\ast}$& $+0.15$ && .....&&&&..... &&&& $-0.23\pm0.11$&4&$+0.81$  \\
O I& 8.66& $-0.02\pm0.10$&2&$+0.23$ && $+0.56\pm0.00$& 1&$+0.73$ && $-0.11\pm0.11$& 2& $+0.47$ && $-0.29\pm0.00$& 1& $+0.75$ \\
Na I& 6.17 & $-0.06\pm0.09$& 3&$+0.19$ &&  $+0.41\pm0.01$& 2&$+0.58$ && $-0.68\pm0.13$& 3& $-0.10$ && $-0.87\pm0.00$& 1& $+0.17$ \\
Mg I& 7.53 & $-0.21\pm0.02$& 2&$+0.04$ && $+0.02\pm0.00$& 1&$+0.19$ &&  $-0.57\pm0.08$& 2& $+0.01$ && $-0.75\pm0.02$& 2&
$+0.29$ \\
Al I& 6.37&  $-0.66\pm0.08$& 4& $-0.41$&& .....&&&& $-0.80\pm0.00$&1& $-0.22$ && .....\\
Si I& 7.51 & $-0.28\pm0.09$&9&$-0.03$ & &
 $-0.11\pm0.17$&5&$+0.06$ && $-0.19\pm0.07$&6&$+0.39$ & &
..... \\
S I& 7.14 & $+0.14\pm0.00$&  1& $+0.39$ && .....&&&&  .....&&&&  $-0.74\pm0.02$& 2&$+0.30$ \\
Ca I& 6.31 & $-0.67\pm0.07$& 8&$-0.42$ &&
 $-0.56\pm0.07$& 3&$-0.39$ &&  $-1.12\pm0.07$& 8 & $-0.54$ &&
$-1.33\pm0.17$&4&$-0.29$ \\
Ca II& 6.31& .....&&&& .....&&&& .....&&&& $-1.37\pm0.00$&1&$-0.33$ \\
Sc II& 3.05 & $-0.51\pm 0.04$& 4& $-0.26$ &&
  $-0.53\pm0.02$& 2&$-0.36$ && .....&&&&
 $-1.23\pm0.01$&3&$-0.19$ \\
Ti I& 4.90 & $-0.68\pm0.05$&4&$-0.43$& & $-0.57\pm0.12$&11&$-0.40$ &&  $-0.88\pm0.09$& 9& $-0.30$ & & ..... \\
Ti II& 4.90 & $-0.57\pm0.16$&  4& $-0.32$&& $-0.68\pm0.01$&  2& $-0.51$ &&  $-0.79\pm0.04$& 2& $-0.21$&& $-1.54\pm0.10$& 4&$-0.50$
  \\
Cr I& 5.64 & $-0.38\pm0.10$&4&$-0.13$& & $-0.14\pm0.03$&2&$+0.03$ && $-0.56\pm0.05$& 4&$+0.02$& & $-1.20\pm0.16$&3&$-0.16$  \\
Cr II& 5.64 &  $-0.36\pm0.02$& 3& $-0.11$& & $-0.23\pm0.00$& 1& $-0.06$ && $-0.66\pm0.03$& 2& $-0.08$& & $-1.19\pm0.13$&7&$-0.15$ \\
Mn I& 5.39 &  $-0.38\pm0.04$&3&$-0.13$ &&  $-0.40\pm0.11$&1s$^{\ast}$&$-0.23$ &&  $-0.76\pm0.12$& 6& $-0.18$ && $-1.05\pm0.19$&2&$-0.01$    \\
Fe  & 7.45 & $-0.25$&&&&  $-0.17$ &&&& $-0.58$&&&&  $-1.04$
 \\
Ni  I& 6.23 & $-0.30\pm0.17$&6&$-0.05$& &
  $-0.22\pm0.11$&18&$-0.05$ && $-0.84\pm0.10$& 10& $-0.26$& &$-0.86\pm0.14$&5&$+0.18$ \\
Zn  I& 4.60 & $-0.16\pm0.01$&2&$+0.09$ && $-0.12\pm0.05$&2&$+0.05$ &&  $-0.86\pm0.07$& 2& $-0.28$ &&  $-1.08\pm0.01$& 2 &$-0.04$ \\
Y II& 2.21& $-0.69\pm0.10$&3&$-0.44$ && $-0.66\pm0.09$& 2&$-0.49$ &&  .....&&&& $-1.33\pm0.07$&2&$-0.29$  \\
Zr I& 2.59&.....&&&& .....&&&& $-1.40\pm0.04$&2&$-0.82$ &&  ..... \\
Zr II& 2.59& .....&&&& $-0.70\pm0.00$& 1&$-0.53$&&  .....&&&& $-1.17\pm0.12$&2&$-0.13$ \\
La II& 1.13& .....&&&& .....&&&& .....&&&& $-1.85\pm0.00$&1&$-0.81$ \\
Ce II& 1.58 & $-0.84\pm0.13$&3&$-0.59$ && $-0.36\pm0.05$&3&$-0.19$&& $-1.35\pm0.04$& 3& $-0.77$ && $-1.63\pm0.00$&1&$-0.59$  \\
Nd II& 1.45& $-1.00\pm0.07$&3&$-0.75$ &&  .....&&&& .....&&&& $-1.28\pm0.00$&1&$-0.24$ \\
Sm II& 1.01& $-0.66\pm0.05$&3&$-0.41$ && $-0.50\pm0.12$& 2&$-0.33$&& .....&&&&..... \\
\hline
\end{tabular}
\flushleft$^{\ast}$ {The number of features synthesized for each element has been indicated.}
\end{changemargin}
\end{table*}

\subsubsection{IRAS 19135+3937}

It is a relatively unexplored object,
 SIMBAD gives only IRAS fluxes and V magnitude.
We list the star as RV-like based on its position in the
IRAS two-color diagram$-$ see Figure \ref{loci1}.
 A time-series monitoring of the high-resolution spectra for this star by Gorlova,
 Van Winckel \& Jorissen (2012) indicated radial-velocity variations of the order of 130-140 days.
 Also the H$\alpha$ profile variations were found to correlate with the radial velocity phase indicating
the possibility of a faint companion surrounded by an accretion disk with an outflow.

Our spectrum is not ideal.
All lines appear
to have a red-shifted absorption component.
We have measured a few reasonably clean lines. The lines of N and Si were heavily
distorted indicating unresolved doubling and hence the abundances
 for these elements could not be determined.
The star is metal poor with [Fe/H] of $-$1.0. The  $\alpha$ elements
 Mg and S yields [$\alpha$/Fe] of $+$0.3 which is expected at this metallicity.
The other $\alpha$ elements Ca and Ti and the s-process elements have negative [X/H]
values which defy simple explanation. As explained in Paper-1, the non-LTE
 correction for Ca I for this temperature and metallicity range could be
 in $+$ 0.1 to 0.2 dex range.
 It is difficult to ascribe this to DG effect since Sc (element  with even higher T$_{c}$)
 is less deficient. Further, the expected enrichment of [Zn/Fe] in DG objects is not seen.

 But IRAS 19135+3937 despite having
 suitable temperature, metallicity  and being
 possible member of a binary (Gorlova, Van Winckel \& Jorissen 2012),
does not show clear signature of DG winnowing as evident from abundances in Table 5.
An overabundance of C\,{\sc i} ([C/Fe]=$+$0.8 dex) is derived from the
measurement of four reasonably clean (but asymmetric) lines.
This star deserves further analysis with spectra
at more stable phases.

\subsection{Extension of the existing analyses} 

  To further our understanding of these objects, 
 we have undertaken  
a contemporary analysis using new grid of model atmospheres,  more accurate
{\itshape log~gf}  and covering  more elements.

\subsubsection{IRAS 01427+4633}

 This was considered as a RV Tauri like object based upon its location in
 two color IRAS diagram and no photometry is available.
  Recently Gorlova et al. (2012)
 group conducted the radial velocity monitoring of this object for nearly three years
 starting 2009 and also conducted abundance analysis.
 Gorlova et al. found this object to be a spectroscopy binary with an orbital period
 of 140.8 $\pm$ 0.2 days and eccentricity $e$ of 0.083 $\pm$0.002 and a$\sin$i of 0.31 AU.

 The model parameters derived from our spectra are presented in Table 2.
 We estimate model atmospheric parameters {\itshape T$_{\rm eff}$, log g},
 and $\xi$  of (6500,0.5,3.5) while
 Gorlova et al. estimate (6250,1.5,4.0).
 Ionization equilibrium is satisfactorily accounted for
 in that $\Delta$ = [X$_{II}$/H] $-$ [X$_{I}$/H]
 is $-0.10$, $-0.06$, $+0.01$ and $+0.01$ for Fe, Si, Ti and Cr respectively as
can be seen in Table 6.

 The star is metal poor with [Fe/H] of $-$0.79 and is enriched in the
$\alpha$ elements Mg, Si, Ca and Ti with an average [$\alpha$/Fe] of $+$0.34.
The low metallicity and the $\alpha$ abundances point towards thick disk membership.
Enrichment of the light elements C, N and O suggests
that the star has
evolved beyond the Red Giant Branch (RGB) but substantial exposure to thermal pulses on the AGB is
unlikely to have occurred as the C/O ratio is around 0.5;
the $s$-process elements are not overabundant.

The abundance analysis by Gorlova et al. (2012) use solar {\itshape gf} values
normalised to solar abundances of Grevesse, Noels \& Sauval (1996) and model atmosphere by
Kurucz (1992) while we use solar abundances from
 Asplund, Grevesse \& Sauval (2005). We therefore chose to make comparison using log$\epsilon$
  derived from the two studies.
 We find difference between Gorlova et al - Present work as follows;
 $+$0.27 for C, $+$0.07 for N,  $-$0.11 for Mg,
 $+$0.10 for Si, $-$0.05 for S, $-$0.15 for Ca, $-$0.26 for Sc,
 $-$0.26 for Ti, $+$0.06 for Cr, $+$0.04 for Mn, $-$0.31 for Ni,
 $-$0.01 for Zn, $-$0.15 for Y and $-$0.13 for Ba.
 For majority of elements the agreement is within $\pm$0.20 dex.

 We did not find very convincing evidence for DG winnowing.
 Although [S/Fe] is $+$0.37, it could be attributed to the fact that
 it is an $\alpha$ element. In fact other $\alpha$ elements like
 Si, Ti also show similar enrichment although [Ca/Fe] is relatively smaller.
 With [Zn/Fe] of $-$0.08 and positive [Sc/Fe], it is very unlikely that
  DG winnowing has affected this object.
  This star  is similar to IRAS 07140-2321 studied in Paper-1 
  in the sense that despite favourable conditions such as warm temperature, presence of
  binary companion and circumstellar material 
  the effect of DG winnowing is not seen thereby highlighting our inadequate understanding of this effect.

\subsubsection{HD 52961}

HD 52961 is a high galactic latitude F type supergiant with a strong IR 
excess and is a semi-regular (SRD) variable with a pulsation period of 70 days (Waelkens et al. 1991). The light
curve exhibits the long-term RVb like phenomenon
in the mean magnitude (seen in RV Tauri stars) thought to be caused
due to the variable circumstellar extinction during orbital motion (Van Winckel et al. 1999).
But the light curve does not exhibit alternate deep and shallow minima typically seen in RV Tauri stars and has been labeled as RV Tauri like object in Figure \ref {loci1}.
Radial velocity monitoring was carried out by Van Winckel et al. (1999) who report
the star
to be a spectroscopic binary with an orbital period of 1310 days.

Deroo et al. (2006) who resolved the dusty disk of HD 52961 using N-band
interferometry
found that the dust emission at 8 $\micron$ originates from a compact
region of diameter 50 AU. Gielen et al. (2009) analyzed a
high resolution IR spectrum and found that it was dominated by spectral features
from both amorphous and crystalline silicates. The high crystallinity fraction and
large sized grains show that the dust grains are strongly processed
and that the disk is long-lived.
Even though the analysis of HD 52961 showing significant
depletion of refractory elements in its photosphere was conducted (Waelkens et al. 1991;
Van Winckel 1995),
an extensive abundance analysis with the most recent {\it gf} values was lacking.
\begin{figure}
\begin{center}
\includegraphics[width=\columnwidth]{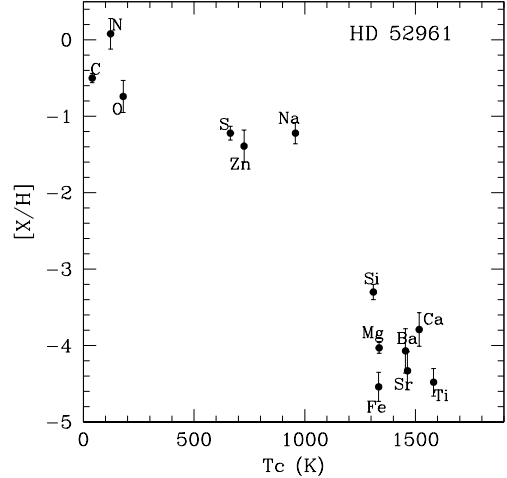}
\caption{Plot of [X/H] versus T$_C$ for HD 52961}
\label{loci5}
\end{center}
\end{figure}
Hence we have performed a detailed abundance analysis for HD 52961 (see Table 6)
using our selection of {\itshape gf} values and Kurucz model atmospheres.
We have compared our results with those of Van
 Winckel (1995); agreement is
within $\pm$0.2 dex for elements in common. We were able to determine for the
first time the abundance of Na, Mg, Si, Ca and Ti as can be seen in Table 6.
 It has resulted in better definition of the depletion curve.
Figure \ref {loci5} shows the plot of [X/H] versus T$_C$ for HD 52961, a classic case of
DG winnowing.

\subsubsection{V453 Oph}

 This is a RV Tauri variable of spectroscopic type RVC and
 photometric type RVa. A photometric
 period of 80 days was obtained by Pollard et al.
 (1996). The star does not have an IR excess (Deroo et al.
2005).

Our spectrum gives
model parameters
($T_{\rm eff} = 5750$ K, $\log$ g =1.50, and $\xi_{\rm t} = 3.7$
km s$^{-1}$) similar to those derived by Deroo et al.
($T_{\rm eff} = 6250$ K, $\log$ g =1.50, and $\xi_{\rm t} = 3.0$
km s$^{-1}$). The star is variable so exact agreement for atmospheric
parameters is not expected. It would appear that we observed the
star at a slightly cooler phase than Deroo et al.
The abundances are in good agreement. 
We find $\delta$ [X/Fe] (present work $-$ Deroo et al.) of $+$0.13 for C, $-$0.12 for O, $+$0.06 for Mg, $+$0.15 for Al, $+$0.18 for Si, $-$0.04 for Ca, $-$0.12 for Sc, $+$0.04 for Ti, $-$0.10 for Cr,
$+$0.30 for Mn, $-$0.13 for Ni, $+$0.23 for Zn,
$+$0.06 for Y, $+$0.07 for Zr, $-$0.06 for Ba, $+$0.23 for La,
$+$0.19 for Ce, $-$0.02 for Nd, $+$0.32 for Eu and $+$0.15 for Dy.
We could determine the additional elements Na and the s-process
element Gd for this star.
But the N abundance could not be determined as our spectrum was obtained
at a cooler phase than that of Deroo et al. (2005).

\begin{table*}[t!]
\begin{changemargin}{1cm}{1cm}
\caption{Elemental Abundances for IRAS 01427+4633, HD 52961 and V453 Oph}
\setlength{\tabcolsep}{0.5\tabcolsep}
\label{table6}
\begin{tabular}{lllllllllllll}
\hline
&& \multicolumn{3}{c}{IRAS 01427+4633}&
\multicolumn{3}{r}{HD 52961} && \multicolumn{3}{r}{V453 Oph}\\
\cline{3-13}  
\multicolumn{1}{l}{Species}&
\multicolumn{1}{l}{$\log \epsilon_{\odot}$}&
\multicolumn{1}{c}{[X/H]}&
\multicolumn{1}{l}{N}&
\multicolumn{1}{l}{[X/Fe]}&
&\multicolumn{1}{c}{[X/H]}&
\multicolumn{1}{l}{N}&
\multicolumn{1}{l}{[X/Fe]}
&& \multicolumn{1}{c}{[X/H]}&
\multicolumn{1}{l}{N}&
\multicolumn{1}{l}{[X/Fe]}
 \\
\hline
C I& 8.39 & $-0.55\pm 0.11$& 5& $+0.24$ && $-0.15\pm0.22$&19&$+4.39$ &&
$-2.39\pm 0.01$& CHs$^{\dagger}$& $-0.13$\\
N I& 7.78 & $+0.02\pm 0.05$& 3& $+0.81$ && $+0.08\pm0.00$& 1&$ +4.62$&& ... \\
O I& 8.66 & $-0.50\pm 0.01$& 3s$^{\ast}$& $+0.29$ &&
$-0.38\pm0.03$& 2& $+4.16$
&&  $-1.39\pm 0.00$& 1& $+0.87$\\
Na I& 6.17 & ...&&&& $-1.10\pm0.01$& 2& $+3.44$
 && $-1.91\pm0.00$& 1&$+0.35$  \\
Mg I& 7.53 & $-0.50\pm 0.06$& 4& $+0.29$&&  $-3.74\pm0.02$& 2&
$+0.80$
&& $-1.95\pm 0.06$& 4& $+0.31$ \\
Mg II& 7.53& ...&&&& $-3.68\pm0.00$& 1&$ +0.86$ && ... \\
Al I & 6.37& ...&&&&...&&&& $-2.68\pm0.04$& 2&$-0.44$\\
Si I& 7.51 & $-0.32\pm0.09$&3&$+0.47$ && ...
 &&&& $-1.61\pm0.06$&2&$+0.65$ \\
Si II& 7.51 & $-0.38\pm 0.00$& 1& $+0.41$&& $-3.08\pm0.00$& 1&$+1.46$&&  $-1.72\pm 0.02$& 2& $+0.54$ \\
S I& 7.14 & $-0.42\pm 0.05$& 2& $+0.37$&&
 $-0.92\pm0.11$& 4&$+3.62$&& ... \\
Ca I& 6.31 & $-0.65\pm0.04$& 5&$+0.14$ &&
... &&&&  $-2.12\pm0.04$& 4&$+0.14$ \\
Ca II& 6.31& ...&&&& $-3.61\pm0.00$& 1&$ +0.90$&& ... \\
Sc II& 3.05 & $-0.41\pm 0.07 $& 4& $+0.38$&&
...&&&&  $-2.10\pm 0.01 $& 1s$^{\ast}$& $+0.16$ \\
Ti I& 4.90 & $-0.31\pm0.09$&2&$+0.48$ && ...&&
&&... \\
Ti II& 4.90 & $-0.30\pm 0.02 $& 2& $+0.49$ && $-4.27\pm0.00$& 1&$+0.27$ &&
$-1.93\pm 0.11 $& 10& $+0.33$ \\
Cr I& 5.64 & $-0.91\pm 0.12 $& 2& $-0.12$&&
... &&&& $-2.49\pm 0.08 $& 4& $-0.23$  \\
Cr II& 5.64 &$-0.90\pm 0.11 $& 6& $-0.11$&&
... &&&& $-2.45\pm 0.07 $& 2& $-0.19$ \\
Mn I& 5.39 &$-0.91\pm 0.07 $& 2& $-0.12$ && ...&&
&&  $-2.59\pm 0.00 $& 1s$^{\ast}$& $-0.33$ \\
Fe  & 7.45 & $-0.79$& & &&$-4.54$ &&&& $-2.26 $ && \\
Ni  I& 6.23 & $-0.55\pm 0.12 $& 5& $+0.24$ &&
...&&&& $-2.52\pm 0.10 $& 4& $-0.26$ \\
Zn  I& 4.60 & $-0.87\pm 0.07 $& 2& $-0.08$ &&
 $-1.27\pm0.09$& 3 &$+3.27$ && $-1.90\pm 0.11 $& 2& $+0.36$ \\
Sr II& 2.92& ...&&&& $-4.33\pm0.00$& 1&$+0.21$ && ... \\
Y  II& 2.21 & $-1.07\pm 0.00 $& 1& $-0.28$ && ...&&&
 & $-1.93\pm 0.06 $& 4& $+0.33$ \\
Zr II& 2.59& ...&&&& ... &&&& $-1.58\pm0.11$& 4&$+0.68$ \\
Ba II & 2.17& $-0.64\pm0.02 $& 1s$^{\ast}$& $+0.15$ && $-4.26\pm0.00$& 1 &$+0.28 $ &&  $-1.91\pm0.08$& 2&$+0.35$\\
La II& 1.13 & ...&&&& ...&&&& $-1.54\pm0.00$& 2& $+0.72$ \\
Ce II& 1.58 & $-0.98\pm 0.14$& 3& $-0.19$ && ...&&&&
$-1.61\pm 0.08 $& 4& $+0.65$   \\
Nd II& 1.45 & ...&&&& ... &&&&
  $-1.60\pm0.12$& 4& $+0.66$ \\
Eu II& 0.52 & ...&&&& ....&&&& $-1.22\pm0.00$& 1s$^{\ast}$&$+1.04$ \\
Gd II& 1.14& ...&&&& ...&&&& $-1.23\pm0.00$& 1s$^{\ast}$&$+1.03$ \\
Dy II& 1.14& ...&&&& ...&&&& $-1.43\pm0.07$& 2&$+0.83$ \\

\hline
\end{tabular}
\flushleft$^{\dagger}$ {refers to synthesis of CH bands.}
\flushleft$^{\ast}$ {The number of features synthesized for each element has been indicated.}
\end{changemargin}
\end{table*}

The metal abundances and the high radial velocity
($-$126 kms$^{-1}$) suggest that V453 Oph is a member of the
galactic halo. There is no evidence that DG winnowing has
affected the star.  An interesting feature is the clear $s$-process
enrichment reported first by Deroo et al. The mean enrichment
[$s$/Fe] is about $+$0.6 with possibly a larger enrichment for
the heavy (Ba-Gd) than for the light (Y,Zr) elements as can be seen in Table 6. This enrichment
is presumably the result of  Third Dredge-Up (TDU) in the AGB progenitor
but the lack of a C-enrichment, also reported by Deroo et al., is
puzzling.
(Our C abundance is obtained from CH lines at 4300-4320\AA~.
Deroo et al's C abundance was obtained from the synthesis of C lines around 9070\AA~.)

\section{Discussion}

 \subsection{Chemical compositions of RV Tauri stars}

 From the compilation of the existing data on the chemical compositions of RV Tauri and RV Tauri like objects
presented in Tables 7-8 (presented in on-line version only), 
it can be seen that they can be divided into certain characteristic groups.
A significant fraction are affected by DG winnowing,
another subgroup show normal composition of evolved red giants. 
Although s-process enhancement is not commonly seen for galactic  RV Tauri stars,
 a few RVCs (two) show mild
s-process enhancement. On the other hand, in the LMC, one RV Tauri star showing significant s-process enhancement
has been detected.
 A couple of galactic RV Tauri stars showing abundance peculiarities correlated with their
FIP have also been reported.
 The stars affected by DG and FIP effect are presented in Table 7 while Table 8 lists stars
 with s-process enhancement and those showing normal composition.

The definition of DI and other parameters listed in Tables 7 \& 8 have been given in the footnote of Table 8. 
The stars in each category  have been arranged in the order of decreasing temperatures. Stars
with and without binary detections have been marked as Y and N in Tables 7 and 8. 
We choose to list only binary detections confirmed via
  direct evidence like radial velocity monitoring.   
This has resulted in smaller fraction of binarity compared to other workers 
(De Ruyter et al. 2006; Gielen et al. 2011) where the detections are based on indirect evidence like the presence of
broad near IR excess (hot dust closer to the star indicating possible 
presence of a dusty circumbinary disk) in their SEDs. This number may
change when more detections are reported.

 \subsubsection{RV Tauri objects showing depletion of refractory elements}

 The main signature of DG winnowing is an observed dependence of depletion
 of condensable elements on the predicted T$_{C}$. Since Fe is also found to be depleted by
 this process, it cannot be the true indicator of intrinsic metallicity hence Zn and S
 which are least affected by this process (due to their low T$_{C}$s)
 would serve as metallicity indicators. We have calculated the intrinsic metallicity: [Fe/H]$_{o}$
 by taking the average of [S/H] and [Zn/H]. For two stars [Zn/H] were not
 available hence we have used only [S/H]. The S being $\alpha$ element, the
 intrinsic metallicity could be over estimated for thick disk
 and halo objects; hence for stars with [Zn/H] lower than -0.7 dex,
 a correction of 0.3 dex for S (see e.g. Maas, Giridhar \& Lambert 2007) and 0.1 dex for
 Zn (see Reddy, Lambert \& Allende Prieto 2006) has been applied for
 calculating the [Fe/H]$_{o}$ for these objects. 
  We consider the star to be
 well depleted when [Zn/Fe] is positive, [Sc/Fe] negative, DI $\geq$ 1.0 and are denoted as DG in Table 7.
 Some stars have incomplete signature of DG winnowing in the sense that
 [Sc/Fe] is negative while [Zn/Fe] is normal and  DIs $<$ 1.0. Such objects
are considered as stars for
 which the DG winnowing is mild and are denoted as MDG in Table 7.

   Similar to the post-AGBs \footnote {Here we refer to post-AGBs as the ones whose light curves
 exhibit small amplitude (0.05-0.5 magnitude) irregular pulsations and do not show the
 typical alternating pattern seen in RV Tauri stars which exhibit large amplitude
 (0.5-1.5 magnitude) pulsations (Kiss et al. 2007). From the positions of RV Tauris
 and a limited sample of post-AGBs with known parallaxes in the H-R diagram, it appears
 that post-AGBs might be systematically more evolved than RV Tauris.},
 the boundary conditions for discernible DG
 i.e. minimum effective temperature and intrinsic metallicity
 (as discussed in Paper-1) are also observed for the RV Tauri stars. These limits are 4500K in temperature and 
  [Fe/H]$_{o}$$\sim$$-$1.0 for intrinsic metallicity are not very different from
 those found for post-AGBs (Sumangala Rao, Giridhar and Lambert 2012).
  As explained in Giridhar, Lambert \& Gonzalez (2000),  
  at cooler temperatures a  deeper convective  envelope  dilutes the accreted gas weakening the abundance anomalies.
 Below the above mentioned metallicity limit the       
lower dust mass fraction results in the inability of the radiation pressure on dust grains to cause DG separation.   
  
  The variation in DG index suggest dependence on several other factors such as
(i) the composition of the gas and dust in the reservoir from which the star draws gas and the temperature distribution within the reservoir
 (ii)  rate and degree of mixing of the accreted material with the atmosphere/envelope of the star.

There is a growing consensus that the DG winnowing is most effective in  RV Tauri variables
 that belong to a binary system with a circumbinary dusty disk providing the reservoir from which
 gas is accreted by the RV Tauri star (Maas, Van Winckel \& Lloyd Evans 2005, Van Winckel 2007).
 For DG and MDG galactic RV Tauri and RV Tauri like stars in Table 7, only 11 out of 33 (33$\%$)
 objects have confirmed binary detections (via radial velocity monitoring).
 The orbital parameters are known accurately only for a handful of these objects as the detection of
 binarity is rendered challenging due to the presence of
stellar pulsations; passage of shock waves during certain phases causing line deformation/splitting.
 Also the radial velocity measurements needs to
 be carried out over several pulsation cycles due to their long periods this
 requires very long observing time.
 High spatial resolution interferometry and radial velocity monitoring are required
 to improve our understanding of these objects.

 In the present work we have detected mild DG winnowing effects in
the additional RV Tauri objects: SU Gem and
BT Lac; mildness of DG effects may be ascribed to their low effective temperatures.
However these stars do not have binary detections and hence radial
 velocity monitoring has to be carried out to confirm binarity.

 \subsubsection{RV Tauris with FIP effect}

 While one generally encounters a normal composition, depletion of refractory elements,
  (not so common) mild s-process enrichment among RV Tauri stars, a couple (CE Vir
  and EQ Cas) exhibited
   very peculiar abundance distribution marked by Na deficiency (see Table 7).
   Rao \& Reddy (2005) showed that the observed abundances were related to the
  their FIP (hence the name FIP effect).
  These authors proposed that the effect  resulted from  systematic removal
  of singly ionized species in the photospheres of these stars under the influence of outflowing
  magnetized columns of gas. The atoms with large FIP  were in 
  neutral state at these temperatures hence were
  were unaffected while those with lower FIP were ionized
  and hence were moved away along with the outflow and hence systematically removed.
 We are not aware of any further addition to this group.

 \subsubsection{RV Tauris with s-process enrichment}

The only known galactic RV Tauri object showing s-process enrichment had been V453 Oph
$-$ an RVC object of long period (80 days). In this work, we
    report a mild s-process enhancement in yet another long period (150 days) RVC
    object V820 Cen. Both these objects exhibit
    large radial velocities, are metal-poor (expected for RVC),
 have no detected IR excess and also show the $\alpha$
    enhancement seen in metal-poor stars.   
  
 Both V820 Cen and V453 Oph could possibly be single stars ; neither there is  
 evidence through radial velocity study nor  indirect evidence through 
 the detection of warm circumstellar matter near the star as they have
 no detected IR excess. A long term radial velocity monitoring  over several pulsation cycles
 would be required to resolve the issue.
 Since only 1$\%$ of post-AGB stars are expected to show extrinsic s-process enrichment
 (through mass-transfer caused by an evolved companion) and no RV Tauris with evolved companions have been detected
 so far (Van Aarle et al. 2013; Deroo et al. 2005),
 it is more likely that these RVC objects are genuinely Thermally Pulsing (TP)-AGB stars.

 It appears that RVC objects are promising candidates to analyse s-processing
    in RV Tauris.  The models of the AGB evolution predict that stars in mass range 1.8 to 4.0M$_{\odot}$
would experience TDU and show s-process enhancement (Herwig 2005) but
 at lower metallicity TDU can take place even at lower masses
 (see Tables 2 and 3 of Karakas, Lattanzio \& Pols 2002). These authors have
     also pointed out the influence of mass-loss on the AGB evolution in
    further lowering this mass limit.

     Hence it is possible that the metal-poor environment of RVC
    might allow thermal pulses and efficient TDU even at the RV Tauri like
    lower mass (0.8 to 1.5M$_{\odot}$) see Deroo et al. (2005) and Reyniers et al.
    (2007) for mass estimates.
  This suggestion is not unreasonable since the
    s-process enriched RV Tauri star detected in the metal-poor environment of the LMC
: MACHO 47.2496.8 (studied by Reyniers et al. 2007) 
 exhibit clear indications
    of s-process enhancement. 

     However there is a major difference; the LMC RV Tauri has C/O greater than 1
     and [s/Fe] is quite large $+$1.2 for lighter and $+$2.1 for heavier s-process
 elements. Galactic RV Tauris exhibit very mild s-processing without carbon enrichment (see Table 8).
     Hence a scenario based upon nucleosynthesis could be an oversimplification.

V453 Oph ([Fe/H]=$-$2.1]) and V820 Cen ([Fe/H]=$-$2.2) are more metal-poor
than the  RVC stars studied so far.
 One wonders if there is a metallicity limit below which RVC might show
s-processing?
We have reservations about considering IRAS 06165+3158 in this group for several reasons.
It is a RV Tauri like object with no photometry, although it is metal-poor with moderate
s-process enhancement, the $\alpha$ enrichment is not seen nor does it show high
radial velocity expected of RVC objects.

  \subsubsection{RV Tauris showing normal composition}

 About  one fourth  of galactic RV Tauri (with and without binary detections) and
 RV Tauri like stars exhibit normal composition i.e. they exhibit
 the signature of ISM for heavy elements while light elements may show
 the effect of CN processing (see Table 8). For significantly metal-poor stars the expected $\alpha$ enrichment
 is observed. Binarity is not very common although a small fraction have binary
 detections.
 Their normal compositions  is understandable as most of them belong to the RVA and C spectroscopic classes
 hence do not meet the boundary conditions for DG effects. 
 However, 
 IRAS 19135+3937 and IRAS 01427+4633 (see Table 8) pose a challenge.
 Despite these objects having favorable conditions for depletion like presence of strong IR excess,
 conducive effective temperatures and binarity,
 they do not show the effects of DG winnowing!

 The absence of observed DG effect
 in these objects might be due to two reasons. Firstly, these  objects are all short period binaries having
 orbital periods and a$\sin$i in the range of 127-141 days and 0.2-0.3 AU respectively.
 (The orbital period range and a$\sin$i for depleted (DG) objects are usually
 in the range of 300-1300 days and 0.4-1.6 AU).
 An optimised orbital period range seem to be an
 additional condition for DG to operate.

 Secondly for  these objects  
Gorlova et al. 2012; Gorlova, Van Winckel \& Jorissen 2012 report          
 a strong P-Cygni H$\alpha$ profile
 with blue shifted absorption components indicating large expansion velocities and splitting of strong
 metal lines which is attributed to 
  the ongoing mass-loss induced by the companion may be responsible for
  absence of depletions observed in these stars. But there is no ready explanation
 for BZ Pyx a binary with an IR excess and orbital period of 371 days.
 We have not yet identified all operators / necessary conditions controlling the DG effect.

\section{Summary and Conclusions}

In this paper we have made a detailed abundance analysis using high resolution
spectra of relatively unexplored RV Tauri stars and stars having RV Tauri like colors from its
position in the IRAS two-color diagram.
A more extensive abundance analysis for the stars V453 Oph and HD 52961
were also undertaken. With the study of additional
elements, a better defined depletion curve for HD 52961 is obtained.
 Our findings are summarized below:

{$\bullet$}
A mild suggestion of DG winnowing is seen in the RV Tauri objects SU Gem
and BT Lac.
No abundance peculiarities are observed in IRAS 01427+4633 and IRAS 19135+3937 despite their binarity
 and the existence of circumstellar material surrounding them.

{$\bullet$}
The detection of small but significant s-processing in yet another RVC star:
V820 Cen is very important as it makes them possible candidates to
study nucleosynthesis near the low mass end of the AGB stars.
 However, it is difficult to account for lack of carbon enrichment in these objects.
Mild s-processing is also detected in the star: IRAS 06165+3158 an
object with RV Tauri like IRAS colors.
Further photometric and spectroscopic monitoring is
 required to understand its variable nature.

Optical-Infrared High-Resolution campaigns of selected RV Tauri
 stars over their pulsation periods may help in understanding
  various abundance patterns observed in these fascinating objects.
 Such a study may also unravel the possible role of pulsations/stellar wind in sustaining depletion.
 One cannot exaggerate the importance of photometric and radial velocity monitoring
  towards the increased detection of binary companions.

Continued studies of unexplored galactic and extragalactic RV Tauri stars could be very rewarding as 
 exemplified by the detection of a few depleted and one s-process enriched RV Tauri stars in the LMC.

\section*{Acknowledgments}
 We are thankful to Prof. David L. Lambert for providing the spectra
 of our program stars and for his helpful suggestions. We are also grateful to Mr G. Selvakumar for
 his help with the VBO Spectra. We are also thankful to the referee whose comments also helped in improving this paper.

\begin{table*}
\addtolength{\tabcolsep}{-1pt}
\begin{changemargin}{-1.8cm}{-1.8cm}
\caption{A compilation of RV Tauri and RV Tauri like variables showing \\ DG winnowing and
FIP effect}
\label{table7}
\begin{tabular}{llllllllllllllll}
\hline
\multicolumn{1}{l}{Star}&
\multicolumn{1}{l}{T$_{\rm eff}$}&
\multicolumn{1}{l}{V$^{a}$}&
\multicolumn{1}{l}{S$^{b}$}&
\multicolumn{1}{l}{[S/H]}&
\multicolumn{1}{l}{[Zn/H]}&
\multicolumn{1}{l}{[Sc/H]}&
\multicolumn{1}{l}{[Ca/H]}&
\multicolumn{1}{l}{[Ti/H]}&
\multicolumn{1}{l}{[Fe/H]}&
\multicolumn{1}{l}{[Fe/H]$_{o}$$^{c}$}&
\multicolumn{1}{c}{DI$^{d}$}&
\multicolumn{1}{l}{P$^{e}$}&
\multicolumn{1}{l}{L$^{f}$}&
\multicolumn{1}{l}{Bin$^{g}$} &
\multicolumn{1}{l}{Ref} \\
\hline
\multicolumn{16}{c}{\bf Depleted (DG) RV Tauri Stars}\\
\hline
HP Lyr& 6300& ...& B& $+$0.0& $-$0.3&
$-$2.8&$-$1.9 &$-$2.9 & $-$0.9& $-$0.2&2.8 & 141& 3.8& Y& 2 \\
DY Ori& 6000& a& B& $+$0.2& $-$0.2&
  $-$2.9&$-$2.2 &$-$2.3 & $-$2.2& $+$0.0&2.7 &60& 3.2& N& 2\\
IW Car & 6700& b& B& $+$0.4&
 $-$0.0&  $-$2.1&$-$1.9 & ... & $-$1.0& $+$0.2&2.4 & 68& 3.3& N& 9 \\
AD Aql& 6300& a& B& $-$0.0& $-$0.1&
  $-$1.8&$-$2.2 &$-$2.6 & $-$2.1& $-$0.1&2.2 &65& 3.3& N& 6 \\
AR Pup & 6000& b& B& $+$0.4& ...&
 $-$2.2&$-$1.4 &... &$-$0.9& $+$0.4$?^{i}$&2.2 &75& 3.4& Y& 5 \\
17233-4330$^{\ast}$& 6250& b& B& $+$0.1 & $-$0.2& $-$1.6&$-$1.4 &$-$1.6 & $-$1.0& $-$0.1&1.7 &...& ... & N& 8 \\
UY CMa& 5500& a& B& $-$0.3& $-$0.6&
  $-$2.2&$-$1.6 &$-$2.4 & $-$1.3& $-$0.5&1.7 &114& 3.7& N& 2 \\
CT Ori& 5500& a& B& $-$0.5& $-$0.5&
  $-$2.5&$-$1.8 &$-$2.5 & $-$1.8& $-$0.5&1.7 &136& 3.8& N& 4 \\
SX Cen & 6500& b& B& $-$0.1&
 $-$0.5& $-$1.9&$-$1.5 &$-$1.9 & $-$1.1& $-$0.3&1.6 & 33& 2.8& Y& 7 \\
R Sge& 5750& b& A& $+$0.4& $-$0.2& $-$1.5&$-$0.9 &$-$1.3& $-$0.5& $+$0.1&1.6 & 71& 3.4& N& 5 \\
 UZ Oph& 5000& a& A& $-$0.4& $-$0.7& $-$1.3&$-$1.1 &$-$1.0 & $-$0.7& $-$0.8&1.5 & 87& 3.5& N& 2 \\
 RX Cap& 5800& ...& A& $-$0.6& $-$0.6& $-$1.2 &$-$0.8 &$-$0.6 & $-$0.8& $-$0.6&1.4 &68& 3.3& N& 2 \\
UY Ara& 5500& a& B& $+$0.0& $-$0.3&
 $-$1.7&$-$1.1 &... & $-$1.0& $-$0.2&1.4 &58& 3.2& N& 1\\
 AZ Sgr& 4750& a& A& $-$0.3& ...& $-$1.8&$-$1.8 &$-$1.6 & $-$1.6& $-$0.3$?^{i}$&1.4 & 114& 3.8& N& 2 \\
BZ Sct& 6250& ...& B& $+$0.2& $+$0.0& $-$1.1&$-$0.9 &$-$1.2 & $-$0.8& $+$0.1&1.3 & ...& ...& N& 2 \\
EP Lyr& 5750& b& B& $-$0.6& $-$0.7&
 $-$2.1&$-$1.8 &$-$2.0 & $-$1.8& $-$0.9&1.3 & 83& 3.5& Y& 5 \\
SS Gem& 5500& a& B& $-$0.4&
 $+$0.0& $-$1.9&$-$1.0 &$-$2.0 & $-$0.8& $-$0.2&1.2 &89& 3.5& N& 4 \\
16230-3410$^{\ast}$& 6250& ...& A& $-$0.4 & $-$0.4& $-$2.3& $-$0.7 & $-$1.4& $-$0.7& $-$0.4&1.1 & ...& ...& N& 8 \\
LR Sco& 6250& a& B& $+$0.0 & $+$0.2& $-$1.1&$-$0.2 &$-$0.6 & $-$0.0& $+$0.1&1.1 &104& 3.6& Y& 2 \\
RU Cen & 6000& a& B& $-$0.7&
 $-$1.0& $-$1.9&$-$1.9 &$-$1.9 & $-$1.9& $-$1.1&1.1 &65& 3.3& Y& 7 \\
AC Her & 6000& a& B& $-$0.4& $-$0.9&
 $-$1.7&$-$1.5 &$-$1.6 &$-$1.4& $-$0.9&1.1 & 75& 3.4& Y& 6 \\
 TT Oph& 4800& a& A& $+$0.0& $-$0.7& $-$1.1&$-$1.1 &$-$0.8 & $-$0.8& $-$0.8&1.0 &61& 3.3& N& 1 \\
 V Vul& 4500& a& A& $+$0.6& $-$0.3& $-$0.7 &$-$0.5 &$-$0.2 & $-$0.4& $+$0.1&1.0 &76& 3.6& N& 2 \\
\hline
\multicolumn{16}{c}{\bf Mildly Depleted (MDG) RV Tauri Stars}\\
\hline
 AI Sco& 5300& b& A& $-$0.1& $-$0.6& $-$0.9&$-$0.6 &$-$0.9 & $-$0.7& $-$0.3&0.8 &71& 3.4& N& 2 \\
 TX Oph& 5000& ...& A& $-$0.6& $-$1.2& $-$1.8&$-$1.4 &$-$1.2 & $-$1.2& $-$1.1&0.8 &135& 3.8& N&2 \\
 SU Gem& 5250& b& A& $+$0.1& $-$0.1& $-$0.5&$-$0.7 &$-$0.6 & $-$0.2& $+$0.0&0.7 &50& 3.2& N& 3 \\
 U Mon& 5000& b& A& $-$0.1& $-$0.7& $-$0.9&$-$0.9 &$-$0.7 & $-$0.8& $-$0.6&0.7 &91& 3.6& Y&1 \\
R Sct & 4500& a& A& ...& $-$0.2& $-$1.4&... &$-$0.4 & $-$0.4& $-$0.2&0.7 &147& 4.0&N& 1 \\
17038-4815$^{\ast}$& 4750& a& A& ... &$-$1.2 & $-$2.0&$-$1.4 &$-$1.9 & $-$1.5& $-$1.2&0.6 & ...& ...& Y & 8 \\
BT Lac& 5000& b& A& ...& $-$0.1& $-$0.5&$-$0.6 &$-$0.6 & $-$0.2& $-$0.1&0.5 &41& 3.0&  N& 3 \\
 DY Aql& 4250& ... & A& ...& ...& $-$2.1 &$-$1.2 &... & $-$1.0& ...&... & 131& 4.0& N& 4 \\
\hline
\multicolumn{16}{c}{\bf Depleted (DG) RV Tauri like Stars$^{h}$} \\
\hline
 AF Crt& 5750& RV like& ...& $-$0.5& $-$0.9& $-$4.2&$-$2.7 &... & $-$2.7& $-$0.9&3.9 &32& ...& Y& 15 \\
HD 52961& 6000& RV like& ...& $-$0.9& $-$1.3& ... &... &... & $-$4.5& $-$1.3&3.8 &71& ...& Y& 3 \\
\hline
\multicolumn{16}{c}{\bf Depleted (DG) RV Tauri Stars in the LMC}\\
\hline
82.8405.15$^{\dagger}$& 6000& ...& ...& $-$0.3& $-$0.3& $-$3.1&$-$2.1 &$-$2.8 & $-$2.5& $-$0.3&2.3 & 93&
 3.5& N& 13 \\
79.5501.13$^{\dagger}$& 5750& ...& ...& $-$0.6& $-$0.6& $-$3.0&$-$1.7 &$-$2.8 & $-$1.8& $-$0.6&1.9 &97& 3.6& N& 14 \\
81.8520.15$^{\dagger}$& 6250& ...& ...& ...& $-$1.4& $-$1.9&$-$1.5 &$-$1.9 & $-$1.6& $-$1.4&0.4& 84& 3.5& N& 14 \\
81.9728.14$^{\dagger}$& 5750& ...& ...& ...& $-$1.2& $-$2.0&$-$1.1 &$-$1.8 & $-$1.1& $-$1.2&0.4& 94& 3.6& N& 14 \\
\hline
\multicolumn{16}{c}{\bf FIP affected RV Tauri Stars}\\
\hline
CE Vir& 4250& ... & A& ...& $-$0.7& $-$2.5&$-$1.6 &$-$1.0 & $-$1.0& $-$0.7& ...&67& 3.5& N & 10\\
EQ Cas& 5300& a& B& $-$0.3& $-$0.3& $-$3.1&$-$2.0 &$-$1.3 & $-$0.8& $-$0.3&... &58& 3.3& N & 2 \\
\hline
\end{tabular}
\end{changemargin}
\end{table*}

\begin{table*}
\begin{changemargin}{-1cm}{-1cm}
\caption{A compilation of RV Tauri and RV Tauri like variables showing\\ s-process enrichment and normal compositions}
\label{table8}
\begin{tabular}{lllllllllllll}
\hline
\multicolumn{1}{l}{Star}&
\multicolumn{1}{l}{T$_{\rm eff}$}&
\multicolumn{1}{l}{V$^{a}$}&
\multicolumn{1}{l}{S$^{b}$}&
\multicolumn{1}{l}{[S/H]}&
\multicolumn{1}{l}{[Si/H]}&
\multicolumn{1}{l}{[Ca/H]}&
\multicolumn{1}{l}{[Fe/H]}&
\multicolumn{1}{l}{[s/Fe]$^{j}$}&
\multicolumn{1}{l}{P$^{e}$}&
\multicolumn{1}{l}{L$^{f}$}&
\multicolumn{1}{l}{Bin$^{g}$} &
\multicolumn{1}{l}{Ref} \\
\hline
\multicolumn{13}{c}{\bf s-process Enriched RV Tauri Stars}\\
\hline
 V453 Oph& 5750& a& C& ...& $-$1.7& $-$2.1& $-$2.2& $+$0.7& 81& 3.5& N& 3\\
V820 Cen& 4750& a& C& ...& ...& $-$1.7& $-$1.9& $+$0.4& 150& 3.9& N& 3 \\
\hline
\multicolumn{13}{c}{\bf s-process Enriched RV Tauri Like Star$^{h}$}\\
\hline
06165+3158$^{\ast}$& 4250& RV like& ...& ...& $-$0.9& $-$1.2& $-$0.9& $+$0.4& ...& ...& N& 3 \\
\hline
\multicolumn{13}{c}{\bf s-process Enriched RV Tauri Star in the LMC}\\
\hline
47.2496.8$^{\dagger}$ & 4900& ...& ...& ...& ...& $-$1.9& $-$1.5& $+$1.8& 113& 3.8& N& 12 \\
\hline
\multicolumn{13}{c}{\bf s-process Enriched RV Tauri like Star in the LMC}\\
\hline
J050632.10-714229.8& 6750& ...& ... & $-$1.2& $-$0.6& $-$1.4& $-$1.2& $+$1.2& 49& 3.7 & N& 16 \\
\hline
\multicolumn{13}{c}{\bf Normal Composition RV Tauri Stars}\\
\hline
BZ Pyx& 6500& b& B& $-$0.7& $-$0.6& $-$0.5& $-$0.7& $+$0.2& ...& ...& Y & 8 \\
DS Aqr& 6500& a& C& $-$0.8& $-$0.8& $-$1.0& $-$1.1& $-$0.1& 77& 3.4& N& 1 \\
EN TrA& 6000& ...& B& $-$0.6& $-$0.6& $-$1.0& $-$0.8& ...& 37& 2.9&
Y& 11 \\
BT Lib& 5800& ...& C& $-$0.8& $-$0.7& $-$0.9& $-$1.2& $+$0.5$?^{k}$& 75& 3.4& N& 1 \\
09538-7622$^{\ast}$& 5500& b& A& $-$0.3 &$+$0.0 & $-$0.6& $-$0.6& $-$0.5& ...& ...& N& 8 \\
AR Sgr& 5300& a& A& $-$0.8& $-$0.8& $-$1.4& $-$1.5& $-$0.1& 88& 3.5& N& 2 \\
V360 Cyg&5250& a& C& $-$0.9& $-$0.9& $-$1.3& $-$1.4& $+$0.0& 70& 3.4& N& 6 \\
DF Cyg& 4800& a& A& ...& $+$0.1& $-$0.2& $-$0.0& $-$0.7& 50& 3.2& N& 2 \\
TW Cam& 4800& a& A& $+$0.0& $-$0.1& $-$0.7& $-$0.5& $+$0.1& 87& 3.6&
N& 1 \\
RV Tau& 4500& b& A& ...& $-$0.3& $-$0.5& $-$0.4& $-$0.4& 79& 3.6& N& 1 \\
TX Per& 4250& a& A& ...& $-$0.2& $-$1.1& $-$0.6& $-$0.8& 78& 3.6& N& 3\\
\hline
\multicolumn{13}{c}{\bf Normal Composition RV Tauri Like Stars$^{h}$}\\
\hline
01427+4633$^{\ast}$& 6500& RV like& ...& $-$0.4& $-$0.4& $-$0.7& $-$0.7& $-$0.1& ...& ...& Y& 3 \\
19135+3937$^{\ast}$& 6000& RV like& ...& $-$0.7& ...& $-$1.3& $-$1.0& $-$0.4& ...& ...& Y& 3 \\
\hline
\end{tabular}
\flushleft$^{\dag}${Refers to the RV Tauri stars in the LMC discovered by Alcock et al. (1998) through the MACHO (Massive Compact Halo Object) project.}
\flushleft$^{\ast}${Stands for IRAS name}
\flushleft$^{a}${Variability Type deduced from the behaviour of the light curves. a and b represent the RV Tauri photometric type which has been explained in the introduction part of this paper.}
\flushleft$^{b}${Spectroscopic Type: A, B and C refer to the Preston's spectroscopic type for RV Tauri stars explained in detail in the introduction part of this paper.}
\flushleft$^{c}${[Fe/H]$_{o}$ refers to the initial metallicity and is calculated by taking the average of [S/H] and [Zn/H].}
\flushleft$^{d}${DI refers to Depletion Index given by DI = [S/H]-{([Ca/H]+[Sc/H]+[Ti/H])/3}. In the absence of other elements DI is calculated from S and Fe.}
\flushleft$^{e}${P refers to Pulsational Period in Days.}
\flushleft$^{f}${L refers to log(L/L$_\odot$) and are the luminosities of the stars calculated from the P-L relation given by Alcock et al. (1998).}
\flushleft$^{g}${Bin refers to binarity. The direct evidence for binarity comes from radial velocity monitoring (Y) while the other stars (N) do not have binary detections as of now.}
\flushleft$^{h}${The RV Tauri like stars do not have published light curves and photometry. These have RV Tauri like IR colors in the two-color diagram and have been considered in this study.}
\flushleft$^{i}${Uncertain [Fe/H]$_{o}$ due to unavailability of [Zn/H].}
\flushleft$^{j}${[s/Fe] corresponds to the average of all the s-process elements}
\flushleft$^{k}${Uncertain abundance as it is based on single line measurement of Ba\,{\sc ii}}
\flushleft$^{1}${Giridhar, Lambert \& Gonzalez 2000}, $^{2}${Giridhar et al. 2005}, $^{3}${Present Work}, $^{4}${Gonzalez, Lambert \& Giridhar 1997a}
\flushleft$^{5}${Gonzalez, Lambert \& Giridhar 1997b}, $^{6}${Giridhar, Lambert \& Gonzalez 1998}, $^{7}${Maas, Van Winckel \& Waelkens 2002}, $^{8}${Maas, Van Winckel \& Lloyd Evans 2005}, $^{9}${Giridhar, Rao \& Lambert 1994}
\flushleft$^{10}${Rao $\&$ Reddy (2005)}, $^{11}${Van Winckel (1997)}, $^{12}${Reyniers et al. 2007}, $^{13}${Reyniers \& Van Winckel 2007}, $^{14}${Gielen et al. 2009}, $^{15}${Van Winckel et al. 2012}, $^{16}${Van Aarle et al. 2013}
\end{changemargin}
\end{table*}


\begin{thebibliography}
\bibitem{} Alcock, C., et al. 1998, AJ, 115, 1921
\bibitem{} Asplund, M., Grevesse, N., \& Sauval, A. J. 2005, in "Cosmic Abundances as Records of Stellar Evolution and Nucleosynthesis". Eds. Thomas G. Barnes III \& Frank N. Bash, ASP Conf.Ser. 336, 25
\bibitem{} Bond, H, E. 1991, in "Evolution of Stars: the Photospheric Abundance Connection". Eds. G. Michaud \& A. Tutukov, IAU Symposium, 145, 341
\bibitem{} Deroo, P., Reyniers, M., Van Winckel, H., Goriely, S., \& Siess, L.
2005, A\&A, 438, 987
\bibitem{} Deroo, P., et al. 2006, A\&A, 450, 181
\bibitem{} De Ruyter, S., Van Winckel, H., Dominik, C., Waters, L. B. F. M., \&
Dejonghe, H. 2005, A\&A, 435, 161
\bibitem{} De Ruyter, S., et al. 2006, A\&A, 448, 641
\bibitem{} Eggen, O. J. 1986, AJ, 91, 890
\bibitem{} Gielen, C., et al. 2009, A\&A, 503, 843
\bibitem{} Gielen, C., et al. 2011, in "Why Galaxies Care About AGB Stars II: Shining Examples and Common Inhabitants". Eds. F. Kerschbaum, T. Lebzelter \& R. F. Wing, ASP Conf.Ser. 445, 281
\bibitem{} Gielen, C., Van Winckel, H., Min, M., Waters, L. B. F. M., \& Lloyd
Evans, T. 2008, A\&A, 490, 725
\bibitem{} Giridhar, S., Lambert, D. L., \& Gonzalez, G. 2000, ApJ, 531, 521
\bibitem{} Giridhar, S., Rao, N. K., \& Lambert, D. L. 1994, ApJ, 437,
476.
427
\bibitem{} Gorlova, N., et al. 2012, A\&A, 542, 27
\bibitem{} Gorlova, N., Van Winckel, H., \& Jorissen, A. 2012, Baltic Astronomy, 21, 165
\bibitem{} Grevesse, N., Noels, A., \& Sauval, A. J. 1996, In: Cosmic Abundances. Eds. S. S. Holt \& G. Sonneborn, ASP Conf.Ser. 99, 117
\bibitem{} Herwig, F. 2005, ARA\&A, 43, 435
\bibitem{} Joy, A. H. 1952, ApJ, 115, 25
\bibitem{} Jura, M. 1986, ApJ, 309, 732
\bibitem{} Karakas, A. I., Lattanzio, J. C., \& Pols, O. R. 2002, PASA, 19, 515
\bibitem{} Kiss, L. L., Derekas, A., Szab\'{o}, G. M., Bedding, T. R., \& Szabados, L. 2007, \mnras, 375, 1338
\bibitem{} Kukarkin, B. V., Parenago, P. P., \& Kholopov, P. N. 1958, General
Catalogue of Variable Stars, 2nd edn. Academy of Sciences of USSR, Moscow
\bibitem{} Kurucz, R. L. 1992, In: The Stellar Populations of Galaxies. Eds.  B. Barbuy and A. Renzini, IAU Symposium No. 149, 225
\bibitem{} Lewis, B. M., Eder, J., \& Terzian, Y. 1990, ApJ, 362, 634
\bibitem{} Lloyd Evans, T. 1985, MNRAS, 217, 493
\bibitem{} Lloyd Evans, T. 1999, In: Asymptotic Giant Branch Stars. Eds. T. Le Bertre, A. Lebre \& C. Waelkens, IAU Symposium No.191, 453
\bibitem{} Maas, T., Giridhar, S., \& Lambert, D. L. 2007, \apj, 666, 378
\bibitem{} Maas, T., Van Winckel, H., \& Lloyd Evans, T. 2005, A\&A, 429, 297
\bibitem{} Maas, T., Van Winckel, H., \& Waelkens, C. 2002, A\&A, 386, 504
\bibitem{} Mathis, J. S., \& Lamers, H. J. G. L. M. 1992, A\&A, 259, L39
\bibitem{} McWilliam, A. 1998, AJ, 115, 1640.
\bibitem{} Miroshnichenko, A. S., et al. 2007, ApJ, 671, 828
\bibitem{} Mucciarelli, A., Caffau, E., Freytag, B, Hans-G\"{u}nter Ludwig, \&
 Bonifacio, P. 2008, A\&A, 484, 841
\bibitem{} Percy, J. R., Bezuhly, M., Milanowski, M., \& Zsoldos, E. 1997, PASP, 109, 264
\bibitem{} Percy, J. R., \& Coffey, J. 2005, JAAVSO, 33, 193
\bibitem{} Planesas, P., Bujarrabal, V., Le Squeren, A. M., \& Mirabel, I. F. 1991, A\&A, 251, 133
\bibitem{} Pollard, K. R., Cottrell, P. L., Kilmartin, P. M., \& Gilmore, A. C.
1996, MNRAS, 279, 949
\bibitem{} Preston, G. W., Krzeminski, W., Smak, J., \& Williams, J. A. 1963,
 ApJ, 137, 401
\bibitem{} Prochaska, J. X., \& McWilliam, A. 2000, ApJ, 537, 57
\bibitem{} Rao, N. K., \& Reddy, B. E. 2005, MNRAS, 357, 235
\bibitem{} Rao, N. K., Sriram, S., Jayakumar, K., \& Gabriel, F. 2005, JA\&A, 26, 331
\bibitem{} Raveendran, A. V. 1989, MNRAS, 238, 945
\bibitem{} Reddy, B. E., Lambert, D. L., \& Allende Prieto, C. 2006, MNRAS, 367, 1329
\bibitem{} Reyniers, M., et al. 2007, A\&A, 461, 641
\bibitem{} Sumangala Rao, S., Giridhar, S., \& Lambert, D. L. 2012, MNRAS, 419, 1254
\bibitem{} Tempesti, P. 1955, MmSAI, 26, 125
\bibitem{} Tull, R. G., MacQueen, P. J., Sneden, C., \& Lambert, D. L. 1995,
PASP, 107, 251
\bibitem{} Van Aarle, E., Van Winckel, H., De Smedt, K., Kamath, D., \& Wood, P. R. 2013, preprint (arXiv:1304.7103V)
\bibitem{} Van Winckel, H. 1995, Ph.D Thesis, K. U. Leuven
\bibitem{} Van Winckel, H. 2003, ARA\&A, 41, 391
\bibitem{} Van Winckel, H., 2007, Baltic Astronomy, 16, 112
\bibitem{} Van Winckel, H., Hrivnak, B. J., Gorlova, N., Gielen, C., \& Lu, W. 2012, A\&A, 542, A53
\bibitem{} Van Winckel, H., Mathis, J. S., \& Waelkens, C. 1992, Nature, 356, 500
\bibitem{} Van Winckel, H., Waelkens, C., Fernie, J. D., \& Waters, L. B. F. M.
 1999, A\&A, 343, 202
\bibitem{} Van Winckel, H., et al. 1998, A\&A, 336, L17
\bibitem{} Venn, K. A., \& Lambert, D. L. 1990, ApJ, 363, 234
\bibitem{} Waelkens, C., Van Winckel, H., Bogaert, E., \& Trams, N. R. 1991,
A\&A, 251, 495
\bibitem{} Waelkens, C., \& Waters, L. B. F. M. 1993, in "Luminous High-Latitude Supergiants".Ed. D. D. Sasselov, ASP Conf. Ser. 45, 219  
\bibitem{} Wallerstein, G. 2002, PASP, 114, 689
\bibitem{} Waters, L. B. F. M., Trams, N. R., \& Waelkens, C. 1992, A\&A, 262, L37
\bibitem{} Zsoldos, E. 1995, JBAA, 105, 238 
\end{thebibliography}
\end{document}